\documentclass[reprint,amsmath,amssymb,aps,prc,nofootinbib,superscriptaddress]{revtex4-1}
\usepackage{bm}
\usepackage{graphicx}
\usepackage{dcolumn}
\usepackage{color}
\newcommand{\be}{\begin{equation}}
\newcommand{\ee}{\end{equation}}
\newcommand{\bea}{\begin{eqnarray}}
\newcommand{\eea}{\end{eqnarray}}
\newcommand{\bfp}{\mathbf{p}}
\newcommand{\bfr}{\mbox{\boldmath $r$}}

\newcommand{\bfbq}{\mbox{\boldmath $Q$}}
\newcommand{\bfz}{\mathbf{z}}

\newcommand{\mbss}[1]{_{\mbox{\scriptsize #1}}}

\newcommand{\mbsu}[1]{\mbox{\scriptsize #1}}

\newcommand{\vphu}{\vphantom{*}}
\newcommand{\vphd}{\vphantom{1}}

\newcommand{\ve}{\varepsilon}

\begin{document}

\title{Spurious dipole mode in random-phase approximation and in the RPA-based models}

\author{V. Tselyaev}
\affiliation{St. Petersburg State University, St. Petersburg, 199034, Russia}
\email{tselyaev@mail.ru}
\date{\today}

\begin{abstract}
The problems related to the existence of the spurious dipole mode (SDM)
in the self-consistent nuclear-structure models are considered.
A method is formulated that allows to eliminate coupling of the SDM
with the physical modes in the extended random-phase approximation (ERPA) theories,
in particular, in the time blocking approximation (TBA)
which is the model of the ERPA type.
It is shown that the application of this method in the realistic TBA calculations of
the $E1$ excitations gives the results which are very close to the results
of the TBA calculations without using this method if the bare external-field $E1$
operators are replaced by the effective ones.
\end{abstract}


\maketitle

\section{Introduction}
\label{sec:Intr}

The existence of the spurious (or the so-called ghost) modes
is a general feature of the models based on the concept of the mean field.
These modes emerge as a result of the symmetry breaking in the ground state
of the quantum many-particle system generated by the mean field,
and for this reason they are usually associated with the Nambu-Goldstone modes.
The spurious modes should have the zero energy, but in the real calculations
of the excitation spectra this condition is fulfilled only approximately
because of the computational limitations.

In the nuclear structure theory, the spurious dipole mode (SDM) is a consequence
of the breaking of translational symmetry in finite nuclei under the condition
that the underlying exact many-body Hamiltonian is translation-invariant.
The translational symmetry is explicitly broken in the independent-particle model
(IPM) in which the dynamics of the nucleus is fully determined by the single-particle
Hamiltonian including the mean field. In the IPM, the SDM is mixed with the
physical modes and cannot be separated from them.
Note that the breaking of translational symmetry in the IPM takes place
even if the mean field is self-consistent, in particular if it is determined
within the Hartree-Fock approximation \cite{RS1980,Blaizot_1986} or within
the density functional theory (DFT, see, e.g., Refs. \cite{Bender_2003,Drut_2010,Grasso_2019}).

The simplest model in which the SDM emerges in the explicit form is the self-consistent
random-phase approximation (RPA). In this model the SDM has exactly zero energy
and thereby is separated from the physical modes, see, e.g., Refs.
\cite{RS1980,Blaizot_1986,Lane_1980}.
This fact is usually treated as the restoration of the translational symmetry
broken at the IPM level.
Nevertheless, in practice, the separation of the SDM from the physical modes
in the self-consistent RPA is not complete because of two reasons:
(i) there is residual mixing of the SDM with the other modes caused by the approximations
of the numerical solution of the RPA equations, and
(ii) all the modes including the SDM enter the spectral representations of
both the exact and the approximate RPA response functions on an equal footing.
%
This in turn leads to two problems:
(i) explicit extraction of the SDM terms from the RPA response function and
elimination of the residual SDM admixtures from the RPA physical modes,
see \cite{Nakatsukasa_2007,Repko_2019,Kvasil_2019}, and
(ii) elimination of the coupling of the SDM with the physical degrees
of freedom in the beyond-RPA models in which the RPA response functions enter
as the building blocks, see
\cite{Tohyama_2004,Tohyama_2007,Gambacurta_2010,Gambacurta_2011,Mizuyama_2012}.
In the present paper these problems are considered with the use of the methods
developed in Refs. \cite{Tselyaev_2013,Tselyaev_2014}.
The analysis is based on the self-consistent RPA constructed within the DFT.

The paper is organized as follows.
In Sec. \ref{sec:scRPA}, the formalism of the self-consistent RPA is outlined
and the properties of the SDM in the RPA are analyzed.
In Sec. \ref{sec:ERPA}, the problem of the SDM in RPA-based models is considered.
In particular, the method of elimination of the SDM in extended RPA theories
is formulated.
Numerical illustrations of these results are presented in Sec. \ref{sec:res}.
Conclusions are given in the last section.

\section{Spurious modes in the self-consistent RPA}
\label{sec:scRPA}

\subsection{RPA framework}
\label{sec:RPAf}

The RPA is the model which enables one to calculate the energies
and the transition amplitudes of the excited states of the quantum
many-particle system. It is described in many textbooks, see, e.g.,
Refs. \cite{RS1980,Blaizot_1986}.
The main RPA equation can be written in the form
\begin{subequations}
\label{mrpae}
\be
\sum_{34} \Omega^{\mbsu{RPA}}_{12,34}\,z^{n}_{34} =
\omega^{\vphu}_n\,z^{n}_{12}
\label{mrpae1}
\ee
or symbolically
\be
\Omega^{\mbsu{RPA}}_{\vphd} |\,z^{n} \rangle = \omega^{\vphu}_n\,|\,z^{n} \rangle\,.
\label{mrpae2}
\ee
\end{subequations}
Here and in the following the numerical indices ($1,2,3,\ldots$)
stand for the sets of the quantum numbers of some single-particle basis,
$\omega^{\vphu}_n$ is the excitation energy, $z^{n}_{12}$ is the transition amplitude.
The RPA matrix $\Omega^{\mbsu{RPA}}_{12,34}$ has the form
\be
\Omega^{\mbsu{RPA}}_{12,34} =
h^{\vphu}_{13}\,\delta^{\vphu}_{42} -
\delta^{\vphu}_{13}\,h^{\vphu}_{42} + \sum_{56}
M^{\mbsu{RPA}}_{12,56}\,{V}^{\vphu}_{56,34}\,,
\label{orpa}
\ee
where $h^{\vphu}_{12}$ is the single-particle Hamiltonian,
${V}^{\vphu}_{12,34}$ is the amplitude of the residual interaction,
$M^{\mbsu{RPA}}_{12,34}$ is the metric matrix in the RPA.
It is defined as
\be
M^{\mbsu{RPA}}_{12,34} =
\delta^{\vphu}_{13}\,\rho^{\vphu}_{42} -
\rho^{\vphu}_{13}\,\delta^{\vphu}_{42}\,,
\label{mrpa}
\ee
where $\rho^{\vphu}_{12}$ is the single-particle density matrix.
The matrices $\rho$ and $h$ satisfy the equations
\be
\rho^2=\rho\,,\qquad [\,h,\rho\,]=0\,,
\label{speqs}
\ee
which play the role of the single-particle equations of motion.

It is convenient to introduce the single-particle basis that diagonalizes
matrices $\rho$ and $h$:
\be
h^{\vphu}_{12} = \ve^{\vphu}_{1}\delta^{\vphu}_{12}\,,
\qquad
\rho^{\vphu}_{12} = n^{\vphu}_{1}\delta^{\vphu}_{12}
\label{spbas}
\ee
where $n^{\vphu}_{1}$ is the occupation number.
In what follows the indices $p$ and $h$ will be used to label
the single-particle states of the particles ($n^{\vphu}_{p} = 0$)
and holes ($n^{\vphu}_{h} = 1$) in this basis.
The matrices $\Omega^{\mbsu{RPA}}_{12,34}$ and $M^{\mbsu{RPA}}_{12,34}$
act in the one-particle--one-hole ($1p1h$) configuration space.
Correspondingly, the transition amplitudes $z^{n}_{12}$ have only $ph$ and $hp$
components.

In the self-consistent RPA based on the DFT,
the single-particle Hamiltonian and the amplitude of the residual interaction
are deduced from some energy-density functional (EDF) $E[\rho]$
and are determined by the formulas
\be
h^{\vphu}_{12} = \frac{\delta E[\rho]}{\delta\rho^{\vphu}_{21}}\,,
\qquad
{V}^{\vphu}_{12,34} =
\frac{\delta^2 E[\rho]}
{\delta\rho^{\vphu}_{21}\,\delta\rho^{\vphu}_{34}}\,.
\label{frpa}
\ee

\subsection{Elimination of the RPA spurious modes in the general case}
\label{sec:SDMext}

Conventional tool for the description of nuclear excitations
in the quantum many-body theory is the response function formalism.
The response function $R(\omega)$ determines the distribution of
the strength of transitions in the nucleus caused by some external field
represented by the single-particle operator $Q$ according to the formulas
\bea
&&S(E)=-\frac{1}{\pi}\;\mbox{Im}\,\Pi(E+i\Delta)\,,
\label{sfdef}\\
&&\Pi(\omega) = - \langle\,Q\,|\,R(\omega)\,|\,Q\,\rangle\,,
\label{poldef}
\eea
where $S(E)$ is the strength function,
$E$ is an excitation energy, $\Delta$ is a smearing parameter,
and $\Pi(\omega)$ is the (dynamic) polarizability.
In the RPA, the response function is a matrix in the $1p1h$ space
defined as
\be
R^{\mbsu{RPA}}_{\vphd}(\omega) = -
\bigl(\,\omega - \Omega^{\mbsu{RPA}}_{\vphd}\bigr)^{-1}
M^{\mbsu{RPA}}_{\vphd}.
\label{rfdef1}
\ee
The spectral representation of the matrix $R^{\mbsu{RPA}}_{\vphd}(\omega)$
can be written in the form (see \cite{Blaizot_1986,Tselyaev_2013}):
\be
R^{\mbsu{RPA}}_{\vphd}(\omega) =
R^{\mbsu{RPA(phys.)}}(\omega) + R^{\mbsu{RPA(spur.)}}(\omega)\,.
\label{rf2t}
\ee
Here $R^{\,\mbsu{RPA(phys.)}}(\omega)$ represents the ``physical'' part of
the function $R^{\mbsu{RPA}}_{\vphd}(\omega)$ and has the form
\be
R^{\mbsu{RPA(phys.)}}(\omega) =
- {\sum_{n}}^{\,\prime}
\frac{\;\mbox{sgn}(\omega^{\vphu}_{n})\,|\,z^{n} \rangle \langle z^{n}|}
{\omega - \omega^{\vphu}_{\,n}}\,,
\label{rfphys}
\ee
where $\omega^{\vphu}_{\,n}$ and $|\,z^{n} \rangle$ are solutions
of Eq.~(\ref{mrpae2}), symbol ${\sum}^{\,\prime}$ means the sum
over all the ``physical'' modes $n$
(that is the modes with non-zero $\omega^{\vphu}_{\,n}$)
for which the following orthonormalization relation is fulfilled
\be
\langle\,{z}^{n}\,|\,M^{\mbsu{RPA}}_{\vphd}
|\,{z}^{n'} \rangle =
\mbox{sgn}(\omega^{\vphu}_{n})\,\delta^{\vphu}_{n,\,n'}\,.
\label{zmz}
\ee
The function $R^{\,\mbsu{RPA(spur.)}}(\omega)$ represents the ``ghost'' part of the
RPA response function caused by the symmetry breaking.
It consists of two terms having the poles at $\omega = 0$ corresponding to the
spurious modes:
\be
R^{\mbsu{RPA(spur.)}}(\omega) =
- \frac{a^{(\,0,1)}}{\omega}
- \frac{a^{(\,0,2)}}{\omega^2}\,.
\label{rfsb}
\ee
The matrices $a^{(\,0,1)}$ and $a^{(\,0,2)}$ are Hermitian and
satisfy the equations, see \cite{Tselyaev_2013}:
\be
\Omega^{\mbsu{RPA}}_{\vphd}a^{(\,0,1)} = a^{(\,0,2)},\quad
\Omega^{\mbsu{RPA}}_{\vphd}a^{(\,0,2)} = 0\,,
\label{b1b2e1}
\ee
\be
a^{(\,0,1)}M^{\mbsu{RPA}}_{\vphd}\,a^{(\,0,k)} = a^{(\,0,k)},
\quad
k = 1,2,
\label{b1b2e2}
\ee
\be
a^{(\,0,1)} = -\mathfrak{P}\,a^{(\,0,1)*}\mathfrak{P},\quad
a^{(\,0,2)} =  \mathfrak{P}\,a^{(\,0,2)*}\mathfrak{P},
\label{b1b2e3}
\ee
where $\mathfrak{P}$ is the permutation operator acting in the space
of the pairs of the single-particle indices:
$\mathfrak{P}^{\vphu}_{12,34}=\delta^{\vphu}_{14}\delta^{\vphu}_{23}$.
In addition, the following closure relation is fulfilled
\be
a^{(\,0,1)} + {\sum_{n}}^{\,\prime}
\mbox{sgn}(\omega^{\vphu}_{n})\,|\,z^{n} \rangle \langle z^{n}|
= M^{\mbsu{RPA}}_{\vphd}.
\label{rpacr}
\ee

The properties (\ref{b1b2e2}) of the matrices $a^{(\,0,1)}$ and $a^{(\,0,2)}$
enables one to introduce a projection operator (see \cite{Tselyaev_2014}):
\be
P = 1 - a^{(\,0,1)}M^{\mbsu{RPA}}_{\vphd}.
\label{defproj}
\ee
From Eq.~(\ref{b1b2e2}) it follows that
\be
P^2=P,\qquad P\,a^{(\,0,1)}=P\,a^{(\,0,2)}=0\,.
\label{ppbk}
\ee
Equations (\ref{zmz}) and (\ref{rpacr}) yield for the ``physical'' modes:
\be
P\,|\,z^{n} \rangle = |\,z^{n} \rangle.
\label{pzn}
\ee
Thus, from (\ref{rf2t}), (\ref{rfphys}), (\ref{rfsb}),
(\ref{ppbk}), and (\ref{pzn}) we obtain
\be
PR^{\mbsu{RPA}}_{\vphd}(\omega)P^{\dag} = R^{\mbsu{RPA(phys.)}}(\omega)\,,
\label{prfp}
\ee
that solves the problem of the elimination of the spurious modes in the RPA
in the general case if the matrix $a^{(\,0,1)}$ is known.

Equation (\ref{prfp}) can be used in particular
if the external field represented by the operator $Q$ can excite the spurious modes.
In this case the dynamic polarizability in Eq. (\ref{poldef}) is divided into the
``physical'' and ``ghost'' parts analogously to Eq. (\ref{rf2t}).
From Eqs. (\ref{prfp}) and (\ref{defproj}) it follows that the ``physical'' part
of the RPA polarizability can be extracted by means of the equation
\be
\Pi^{\mbsu{RPA(phys.)}}(\omega) = -
\langle\,Q^{\mbsu{eff}}\,|\,R^{\mbsu{RPA}}(\omega)\,|\,Q^{\mbsu{eff}}\,\rangle\,,
\label{polrpa}
\ee
where $Q^{\mbsu{eff}}$ is the effective external-field operator determined as
\be
Q^{\mbsu{eff}} = P^{\dag}Q = \bigl(1 - M^{\mbsu{RPA}}_{\vphd} a^{(\,0,1)}\bigr)\,Q\,.
\label{defqeff}
\ee
The above equation is analogous to the results of Ref. \cite{Repko_2019}
obtained within the framework of the quasiparticle RPA formalism.

\subsection{The case of the SDM}
\label{sec:sdmc}

The explicit form of the matrices $a^{(\,0,1)}$ and $a^{(\,0,2)}$
can be found in the important case of the SDM.
It is known (see Ref.~\cite{RS1980}) that the eigenvectors of this mode
are non-normalizable in the sense of Eq.~(\ref{zmz}).
So, it is convenient to proceed in the way described in Ref. \cite{Tselyaev_2014}.
Let us represent the EDF $E[\rho]$ in Eqs. (\ref{frpa}) as a sum of two terms
\be
E[\rho] = \mbox{Tr}\,\bigl(\rho\,h^0\bigr)
+ E^{\vphu}_{\mbsu{int}}[\rho]\,,
\label{edfdec}
\ee
where $h^0$ is a single-particle operator
and the term $E^{\vphu}_{\mbsu{int}}[\rho]$ contains all contributions
to the total energy related to the interaction.
Actually, $h^0$ is a simple kinetic-energy operator but in order to deal
in the following with the normalizable solutions of the RPA equations
we include an oscillator potential into $h^0$ by setting
\be
h^0 = \frac{\bfp^2}{2m} + \frac{\omega^2_0 m \bfr^2}{2\hbar^2}\,,
\label{h0def}
\ee
where $\bfp=-i\hbar\nabla$ is the momentum operator,
$m$ is the nucleon mass (generally different for the neutrons and protons).
The functional $E^{\vphu}_{\mbsu{int}}[\rho]$ is supposed to be
invariant under the symmetry transformations of the type
\be
E^{\vphu}_{\mbsu{int}}[e^{-i\alpha q}\rho\,e^{i\alpha q}] =
E^{\vphu}_{\mbsu{int}}[\rho]\,,
\label{eintin}
\ee
where $q$ is a Hermitian single-particle operator
and $\alpha$ is an arbitrary real parameter.
Differentiating Eq.~(\ref{eintin}) with respect to $\alpha$ and $\rho$
and setting $\alpha=0$ we obtain
\be
[\,h,q\,]^{\vphu}_{12} +
\sum_{34}V^{\vphu}_{12,34}\,[\,q,\rho\,]^{\vphu}_{34} =
[\,h^0,q\,]^{\vphu}_{12}\,,
\label{hqrho}
\ee
where Eqs. (\ref{frpa}) and (\ref{edfdec}) were taken into account.
Multiplying Eq.~(\ref{hqrho}) from the left with the matrix
$M^{\mbsu{RPA}}$ and using definitions (\ref{orpa}), (\ref{mrpa}),
and equality $[\,h,\rho\,]=0$, we get
\be
\sum_{34}\Omega^{\mbsu{RPA}}_{12,34}\,[\,q,\rho\,]^{\vphu}_{34} =
[\,[\,h^0,q\,],\rho\,]^{\vphu}_{12}\,.
\label{omgqrho}
\ee

Now we note that the functional $E^{\vphu}_{\mbsu{int}}[\rho]$
should be invariant under the translations and the Galilean transformations,
and that it is a common property of all traditional nuclear EDFs,
see \cite{Bender_2003,Drut_2010,Grasso_2019}
(notice, however, that in the general case the Galilean invariance is compatible
with the isotopic symmetry only on the assumption that the masses of neutrons
and protons are equal to each other).
This means that Eqs. (\ref{eintin})--(\ref{omgqrho}) should be
fulfilled for those operators $q$ which are the space components of
the momentum operator $\bfp$
or of the coordinate operator multiplied by the nucleon mass ($m\bfr$).
In the case of the operator $h^0$ defined by Eq.~(\ref{h0def}) we have
\be
[\,h^0,\nabla\,] = - \frac{\omega^2_0}{\hbar^2}\,m\bfr\,,\qquad
[\,h^0,m\bfr\,]  = - \hbar^2\,\nabla\,.
\label{h0com}
\ee
From Eqs. (\ref{omgqrho}) and (\ref{h0com}), after some algebra we
arrive at the following equation
\be
\sum_{34}\Omega^{\mbsu{RPA}}_{12,34}\,\bfz^{(\pm)}_{34} =
\pm\,\omega^{\vphu}_0\,\bfz^{(\pm)}_{12},
\label{rpass}
\ee
where
%
\be
\bfz^{(\pm)}_{12} =
\frac{\hbar}{\sqrt{2\omega^{\vphu}_0 M^{\vphu}_0}}
\biggl( [\,\nabla,\rho\,]^{\vphu}_{12}
\mp \frac{\omega^{\vphu}_0}{\hbar^2}\,
[\,m\bfr,\rho\,]^{\vphu}_{12}\biggr),
\label{zssdef}
\ee
$M^{\vphu}_0=\mbox{Tr}(\rho m)$ is the total mass of the nucleus
and it is supposed that $\omega^{\vphu}_0 > 0$.
The transition amplitudes $\bfz^{(\pm)}_{12}$ are normalized
according to Eq.~(\ref{zmz}).
These amplitudes represent the explicit solutions of the RPA
eigenvalue equation (\ref{mrpae1}) obtained from the symmetry properties
of the EDF.
They correspond to the spurious $1^-$ excitations.
In Ref. \cite{Tselyaev_1998}, the analogous formulas were obtained
for the transition amplitudes of the SDM entering the spectral representation
of the exact response function.

If $\omega^{\vphu}_0$ is finite, we can substitute the solutions (\ref{zssdef})
into the right hand side (r.h.s.) of Eq.~(\ref{rfphys}).
In the limit $\omega^{\vphu}_0 \to +0$ the contribution
of these solutions into the RPA response function takes the form of the
r.h.s. of Eq.~(\ref{rfsb}) with
\bea
&&\hspace{-1em}a^{(\,0,1)}_{12,34} = \frac{1}{M^{\vphu}_0}
\nonumber\\
&&\hspace{-1em}\times \biggl(
[\,\nabla,\rho\,]^{\vphu}_{12}\cdot[\,m\bfr,\rho\,]^{\vphu}_{43}
- [\,m\bfr,\rho\,]^{\vphu}_{12}\cdot[\,\nabla,\rho\,]^{\vphu}_{43}\biggr),
\label{a01e1}\\
&&\hspace{-1em}a^{(\,0,2)}_{12,34} = \frac{\;\,\hbar^2}{M^{\vphu}_0}\,
[\,\nabla,\rho\,]^{\vphu}_{12}\cdot[\,\nabla,\rho\,]^{\vphu}_{43}\,.
\label{a02e1}
\eea
In the particle-hole ({\it p-h}) representation defined by Eqs. (\ref{spbas}),
the formulas (\ref{rf2t}), (\ref{rfphys}), (\ref{rfsb}), (\ref{a01e1}), and (\ref{a02e1})
correspond to Eq. (10.51b) of Ref. \cite{Blaizot_1986}.

The matrices $a^{(\,0,1)}$ and $a^{(\,0,2)}$ defined by
Eqs. (\ref{a01e1}) and (\ref{a02e1}) are Hermitian.
It is not difficult to verify that they satisfy
Eqs. (\ref{b1b2e1})--(\ref{b1b2e3}). Therefore, the projection operator
defined by Eqs. (\ref{defproj}) and (\ref{a01e1})
also satisfies all the conditions described in Sec.~\ref{sec:SDMext}.

Equations (\ref{defqeff}) and (\ref{a01e1}) can be used for determination of
the effective external-field operators in the case of the electric dipole
excitations. Consider the local vector $E1$ operator $Q$ of the form
\be
\bfbq = f_{\tau}(r)\,\bfr\,,
\label{qe1b}
\ee
where $\tau$ is the isotopic index ($\tau = n,p$) and $r=|\bfr|$.
In the following, two kinds of the (bare) radial formfactors $f_{\tau}(r)$
will be considered:
\be
f_{\tau}(r) = \delta_{\tau,p}\,C_{1}\,, \quad C_{1}=e\,\sqrt{3/4\pi}\,,
\label{fe1b}
\ee
in which case $\bfbq$ in Eq.~(\ref{qe1b}) is the usual electric dipole operator,
see \cite{RS1980}, and
\be
f_{\tau}(r) = C_{0}r^2,
\label{fe1isb}
\ee
where $C_{0}$ is a constant, in which case $\bfbq$ is the isoscalar $E1$ operator.
Assuming that the local single-particle density $\rho^{\vphu}_{\tau}(r)$ is spherically symmetric
[$\,\rho^{\vphu}_{\tau}(r)=\sum_{s} \rho(\bfr,s,\tau;\,\bfr,s,\tau)$,
where $s$ is the spin variable],
one obtains from (\ref{defqeff}), (\ref{a01e1}), and (\ref{qe1b})
\be
\bfbq^{\mbsu{eff}} = f^{\mbsu{eff}}_{\tau}(r)\,\bfr\,,
\label{qe1eff}
\ee
where
\be
f^{\mbsu{eff}}_{\tau}(r) = f^{\vphu}_{\tau}(r) -
\frac{m^{\vphu}_{\tau}}{M^{\vphu}_0}\bar{f},
\label{dfeff}
\ee
\be
\bar{f} = \sum_{\tau} \int d\bfr \rho^{\vphu}_{\tau}(r)
\Bigl[f^{\vphu}_{\tau}(r) + \frac{r}{3} f^{\prime}_{\tau}(r)\Bigr].
\label{defbf}
\ee

In the case of the formfactor (\ref{fe1b}), Eqs. (\ref{dfeff}) and (\ref{defbf}) yield
\be
f^{\mbsu{eff}}_{p}(r) = \tilde{C}_1 N/A,\qquad
f^{\mbsu{eff}}_{n}(r) = - \tilde{C}_1 Z/A,
\label{feffiv}
\ee
where $\tilde{C}_1 = C_1 A m^{\vphu}_{n}/M^{\vphu}_0$, $A=N+Z$,
$M^{\vphu}_0 = N m^{\vphu}_{n} + Z m^{\vphu}_{p}$,
$N$ and $Z$ are the numbers of neutrons and protons, respectively.
If in addition the equality $m^{\vphu}_{n}=m^{\vphu}_{p}$ is assumed,
Eqs. (\ref{qe1eff}) and (\ref{feffiv}) correspond to the known formulas
for the effective nucleon $E1$ charges, see \cite{RS1980}.

In the case of the formfactor (\ref{fe1isb}),
assuming again the equality $m^{\vphu}_{n}=m^{\vphu}_{p}$,
one obtains from Eqs. (\ref{dfeff}) and (\ref{defbf})
\be
f^{\mbsu{eff}}_{\tau}(r) = C_{0}
\Bigl(r^2 - \frac{5}{3}\langle r^2 \rangle \Bigr)\,,
\label{feffis}
\ee
where $\langle r^2 \rangle = \mbox{Tr}(\rho r^2)/A$,
that also corresponds to the known formula \cite{Giai_1981} for this case.

Further, from Eq. (\ref{zmz}) it follows that the spurious amplitudes (\ref{zssdef})
at the finite $\omega^{\vphu}_0$
are orthogonal to the ``physical'' ones $|\,z^{n} \rangle$, that is
\be
\langle\,{z}^{n}\,|\,M^{\mbsu{RPA}}_{\vphd}
|\,\bfz^{(\pm)} \rangle = 0\,.
\label{zmzs}
\ee
Using this equality and the definitions (\ref{mrpa}) and (\ref{zssdef})
we obtain
\be
\langle\,{z}^{n}\,|\,m\bfr \rangle = 0\,,\qquad
\langle\,{z}^{n}\,|\,\nabla \rangle = 0\,.
\label{zszer}
\ee
Obviously, Eqs. (\ref{zszer}) remain valid in the limit $\omega^{\vphu}_0 \to +0$.
From Eqs. (\ref{a01e1}) and (\ref{a02e1}) we also have
\bea
a^{(\,0,1)}\,|\,m\bfr \rangle &=& |\,[\,m\bfr,\rho]\,\rangle,
\label{a01mr}\\
a^{(\,0,2)}\,|\,m\bfr \rangle &=& - \hbar^2 |\,[\,\nabla,\rho]\,\rangle\,,
\label{a02mr}
\eea
\bea
a^{(\,0,1)}\,|\,\nabla \rangle &=& |\,[\,\nabla,\rho]\,\rangle,
\label{a01nb}\\
a^{(\,0,2)}\,|\,\nabla \rangle &=& 0\,.
\label{a02nb}
\eea

Note that in the derivation considered above, only the symmetry properties
of the EDF $E[\rho]$ entering the RPA equations (\ref{frpa}) were used.
The symmetry properties of the underlying exact many-body Hamiltonian
formally were not taken into account, though they should coincide
with the properties of the EDF. Such an approach is most adequate
to the framework of the self-consistent RPA.

\section{SDM in RPA-based models}
\label{sec:ERPA}

\subsection{RPA treatment of core-polarization effects in even-odd nuclei
and the SDM}
\label{sec:eon}

Consider the problem of description of the transition probabilities
for the states of the even-odd nuclei supposing that these states
are single-particle and that the transitions between them
are caused by the external field represented by the single-particle operator $Q$.
In this case the transition probabilities are determined by the
matrix elements $\tilde{Q}_{12}$ of the effective operator $\tilde{Q}$
which is the sum of the bare operator $Q$ and the core-polarization term
$\Delta Q$:
\be
\tilde{Q}_{12} = Q_{12} + \Delta Q_{12}\,.
\label{def:tq}
\ee
The treatment of the term $\Delta Q$ within the RPA was
developed and described in Refs. \cite{Migdal_1967,Ring_1973a,Speth_1977}
(see also discussion of this treatment in \cite{Tselyaev_2022}).
It can be represented in the form
\be
\Delta Q^{\vphu}_{12} = - \sum_{3456} {V}^{\vphu}_{12,34}\,
R^{\mbsu{RPA}}_{34,56}(\ve^{\vphu}_{12})\,Q^{\vphu}_{56}\,,
\label{def:dq}
\ee
where the RPA response function $R^{\mbsu{RPA}}_{\vphd}(\omega)$
is defined in Eq. (\ref{rfdef1}),
$\ve^{\vphu}_{12} = \ve^{\vphu}_{1} - \ve^{\vphu}_{2}$
with $\ve^{\vphu}_{1}$ determined by Eq. (\ref{spbas}).

If the external field is represented by the operators
$\bfbq=m\bfr$ or $\bfbq=\nabla$, it acts only on the center-of-mass coordinate
and thus should not give rise to the transitions between the intrinsic
states of the nucleus. So, in this case the effective operator $\tilde{\bfbq}$
should be equal to zero.

As follows from the results of Sec.~\ref{sec:sdmc},
in the self-consistent RPA from Eqs. (\ref{rfphys}) and (\ref{zszer}) we have
%
%
\be
R^{\mbsu{RPA(phys.)}}_{\vphd}(\omega)\,|\,m\bfr \rangle\,=\,
R^{\mbsu{RPA(phys.)}}_{\vphd}(\omega)\,|\,\nabla \rangle\,=\,0\,.
\label{rfphyscm}
\ee
On the other hand, from Eqs. (\ref{spbas}), (\ref{rfsb}),
(\ref{hqrho}), (\ref{h0com}), (\ref{a01mr})--(\ref{a02nb})
it follows that
\bea
&&\sum_{3456}
{V}^{\vphu}_{12,34}\,R^{\mbsu{RPA(spur.)}}_{34,56}(\ve^{\vphu}_{12})\,
(m\bfr)^{\vphu}_{56} = (m\bfr)^{\vphu}_{12}\,,
\label{vrspmr}\\
&&\sum_{3456}
{V}^{\vphu}_{12,34}\,R^{\mbsu{RPA(spur.)}}_{34,56}(\ve^{\vphu}_{12})\,
\nabla^{\vphu}_{56} = \nabla^{\vphu}_{12}\,.
\label{vrspnb}
\eea
Therefore, from Eqs. (\ref{rf2t}), (\ref{def:tq})--(\ref{vrspnb})
we obtain that $\tilde{\bfbq}=0$ for the external-field operators
$\bfbq=m\bfr$ and $\bfbq=\nabla$ if the quantities $V$,
$R^{\mbsu{RPA}}_{\vphd}(\omega)$, and $\ve^{\vphu}_{12}$
in (\ref{def:dq}) are determined in the self-consistent RPA
and if the contribution of the SDM is included in the
$R^{\mbsu{RPA}}_{\vphd}(\omega)$.

\subsection{Elimination of the SDM in extended RPA theories}
\label{sec:erpa}

It is well known that in the general case the response function
can be defined as a solution of the Bethe-Salpeter equation
(BSE, see, e.g., Ref.~\cite{Speth_1977}).
In the case of the RPA response function defined by Eq. (\ref{rfdef1}),
the BSE reads
\be
R^{\mbsu{RPA}}_{\vphd}(\omega) = R^{(0)}_{\vphd}(\omega)
- R^{(0)}_{\vphd}(\omega)VR^{\mbsu{RPA}}_{\vphd}(\omega)\,,
\label{rfrpa}
\ee
where $R^{(0)}_{\vphd}(\omega)$ is uncorrelated {\it p-h}
propagator and $V$ is the amplitude of the residual interaction
entering the RPA matrix (\ref{orpa}). All the matrices in Eq.~(\ref{rfrpa})
are defined in the $1p1h$ configuration space.
The {\it p-h} propagator $R^{(0)}_{\vphd}(\omega)$ is defined as
\be
R^{(0)}_{\vphd}(\omega) = -
\bigl(\,\omega - \Omega^{(0)}_{\vphd}\bigr)^{-1}
M^{\mbsu{RPA}}_{\vphd},
\label{rf0}
\ee
where
\be
\Omega^{(0)}_{12,34} =
h^{\vphu}_{13}\,\delta^{\vphu}_{42} -
\delta^{\vphu}_{13}\,h^{\vphu}_{42}
\label{omrpaz}
\ee
and $M^{\mbsu{RPA}}_{\vphd}$ is the metric matrix (\ref{mrpa}).

In the beyond-RPA models, the $1p1h$ configuration space of the RPA is extended
by including more complex configurations, e.g. of the $2p2h$, $1p1h\otimes$phonon
or two-phonon type, where the phonons in the simplest case are superpositions
of the $1p1h$ configurations represented by the solutions
of the RPA equation (\ref{mrpae}).
In the following we consider extended RPA (ERPA) theories in which
the response function is determined by the BSE of the form
\bea
R^{\mbsu{ERPA}}_{\vphd}(\omega) &=& R^{(0)}_{\vphd}(\omega)
\nonumber\\
&-& R^{(0)}_{\vphd}(\omega)\,[V + \bar{W}(\omega)]\,R^{\mbsu{ERPA}}_{\vphd}(\omega)
\label{rferpa1}
\eea
which can be also rewritten as
\bea
R^{\mbsu{ERPA}}_{\vphd}(\omega) &=& R^{\mbsu{RPA}}_{\vphd}(\omega)
\nonumber\\
&-& R^{\mbsu{RPA}}_{\vphd}(\omega) \bar{W}(\omega) R^{\mbsu{ERPA}}_{\vphd}(\omega)\,.
\label{rferpa2}
\eea
Here $\bar{W}(\omega)$ is the (subtracted) amplitude of the induced interaction
including contributions of complex configurations. This amplitude
is the energy-dependent matrix in the $1p1h$ space. It has the form
\be
\bar{W}^{\vphu}_{12,34}(\omega) = W^{\vphu}_{12,34}(\omega) - W^{\vphu}_{12,34}(0)
\label{def:bw}
\ee
with
\be
{W}^{\vphu}_{12,34}(\omega) =
\sum_{c,\;\sigma}\,\frac{\sigma\,
{F}^{c(\sigma)}_{12}
{F}^{c(\sigma)*}_{34}}
{\omega - \sigma\,\Omega^{\vphu}_{c}}\,,
\label{def:w}
\ee
where $\sigma = \pm 1$ and $\,c$ is an index of the subspace of
complex configurations. Explicit formulas for the amplitudes
${F}^{c(\sigma)}_{12}$ and the energies $\Omega^{\vphu}_{c}$
in the case of some models of the ERPA type are given in Ref.~\cite{Tselyaev_2013}.
Note that in the models considered in \cite{Tselyaev_2013} all the energies
of complex configurations $\Omega^{\vphu}_{c}$ are positive.

The subtraction of the amplitude $W(0)$ in Eq. (\ref{def:bw}) ensures
the stability of solutions of the ERPA equations (see \cite{Tselyaev_2013}
for more details).
At present, this subtraction method is used in the beyond-RPA models including
$1p1h\otimes$phonon
\cite{Litvinova_2007,Lyutorovich_2015,Lyutorovich_2016,Tselyaev_2016,Niu_2016,Roca-Maza_2017}
and $2p2h$ \cite{Gambacurta_2015,Gambacurta_2016} configurations.
As follows from Eq. (\ref{rferpa2}), the subtraction of $W(0)$ also ensures
the existence of the poles of $R^{\mbsu{ERPA}}_{\vphd}(\omega)$ at $\omega = 0$
if these poles (corresponding to the spurious modes) are presented in the
RPA response function determined by Eq. (\ref{rfrpa}).
This property was one of the motivations for introducing this method
in Ref.~\cite{Tselyaev_2007}.
However, the subtraction method in itself does not exclude the residual coupling
of the spurious modes with the physical ones mediated by the amplitude
$\bar{W}(\omega)$ at the non-zero energies.
As a result, though the main component of the spurious state appears
at zero energy, its fragments can be spread out over the wide energy range.

Formally, the contribution of the spurious states is completely absent
in the response function ${R}^{\mbsu{ERPA(phys.)}}_{\vphd}(\omega)$
defined by the equation
\bea
&&\hspace{-4em} R^{\mbsu{ERPA(phys.)}}_{\vphd}(\omega) =
R^{\mbsu{RPA(phys.)}}_{\vphd}(\omega)
\nonumber\\
&-& R^{\mbsu{RPA(phys.)}}_{\vphd}(\omega) \bar{W}(\omega)
R^{\mbsu{ERPA(phys.)}}_{\vphd}(\omega)\,,
\label{rfbperpa}
\eea
where the function $R^{\mbsu{RPA(phys.)}}_{\vphd}(\omega)$
is formally determined by Eq. (\ref{rfphys}).
In the case of the SDM in the self-consistent RPA,
the function $R^{\mbsu{RPA(phys.)}}_{\vphd}(\omega)$ can be in practice determined
by Eqs. (\ref{prfp}), (\ref{defproj}) and (\ref{a01e1}).
However, such determination of ${R}^{\mbsu{ERPA(phys.)}}_{\vphd}(\omega)$ requires
preliminary solving the RPA BSE (\ref{rfrpa}) that complicates the task.

The more convenient method of the elimination of the SDM in the ERPA
was suggested in Ref. \cite{Tselyaev_2014}.
Consider the function ${R}^{\mbsu{ERPA+}}_{\vphd}(\omega)$
which is a solution of the equation
\bea
\hspace{-2em}{R}^{\mbsu{ERPA+}}_{\vphd}(\omega) &=& R^{(0)}_{\vphd}(\omega)
\nonumber\\
&-& R^{(0)}_{\vphd}(\omega)\,[V + \bar{W}^{\perp}(\omega)]\,
{R}^{\mbsu{ERPA+}}_{\vphd}(\omega)\,,
\label{rfberpa1}
\eea
where the amplitude $\bar{W}^{\perp}(\omega)$ is defined as
\be
\bar{W}^{\perp}(\omega)=P^{\dag}\bar{W}(\omega)P
\label{wperp}
\ee
with the operator $P$ defined in Eqs. (\ref{defproj}) and (\ref{a01e1}).
In fact, Eq. (\ref{rfberpa1}) is analogous to Eq. (\ref{rferpa1}) and is equivalent
to Eq. (\ref{rferpa2}) in which the amplitude
$\bar{W}(\omega)$ is replaced by $\bar{W}^{\perp}(\omega)$.
But as follows from Eqs. (\ref{rfsb}) and (\ref{ppbk}), the amplitude $\bar{W}^{\perp}(\omega)$
is orthogonal to the ``ghost'' part of the RPA response function,
because
\be
P R^{\mbsu{RPA(spur.)}}_{\vphd}(\omega) = R^{\mbsu{RPA(spur.)}}_{\vphd}(\omega) P^{\dag} = 0\,.
\ee
So, the coupling of the spurious modes with the physical ones in Eq. (\ref{rfberpa1})
is eliminated.

After a series of transformations, one can show that the solution of
Eq. (\ref{rfberpa1}) satisfies the equality
\be
{R}^{\mbsu{ERPA+}}_{\vphd}(\omega) =
{R}^{\mbsu{ERPA(phys.)}}_{\vphd}(\omega) +
{R}^{\mbsu{RPA(spur.)}}_{\vphd}(\omega)\,,
\label{rfberpa2}
\ee
where ${R}^{\mbsu{ERPA(phys.)}}_{\vphd}(\omega)$ and
${R}^{\mbsu{RPA(spur.)}}_{\vphd}(\omega)$ are defined by
Eqs. (\ref{rfbperpa}), (\ref{rfsb}), (\ref{a01e1}), and (\ref{a02e1}).
Equation (\ref{rfberpa2}) explicitly shows that the SDM
is fully separated from the physical modes in the response function
${R}^{\mbsu{ERPA+}}_{\vphd}(\omega)$.
At the same time, this function is determined by only one Eq. (\ref{rfberpa1})
which is the main equation of the given method.
In what follows the model corresponding to this equation will be referred to
as the ERPA with projection (ERPA+).

The problem of extracting the ``physical'' part of the RPA response function
in the beyond mean-field model was considered in Ref. \cite{Mizuyama_2012}.
Within the formalism described above, the method of Ref. \cite{Mizuyama_2012}
corresponds to introducing the corrected RPA response function
$\tilde{R}^{\mbsu{RPA}}_{\vphd}(\omega) = \tilde{P} R^{\mbsu{RPA}}_{\vphd}(\omega)$
where
\be
\tilde{P} = 1 - \tilde{a}^{(\,0,1)}M^{\mbsu{RPA}}_{\vphd},
\label{deftproj}
\ee
\be
\tilde{a}^{(\,0,1)}_{12,34} = \frac{1}{M^{\vphu}_0}\,
[\,\nabla,\rho\,]^{\vphu}_{12}\cdot[\,m\bfr,\rho\,]^{\vphu}_{43}\,.
\label{ta01e1}
\ee
Comparing Eqs. (\ref{deftproj}) and (\ref{ta01e1}) with
Eqs. (\ref{defproj}) and (\ref{a01e1}) for the operator $P$, one can see that
\be
P = \tilde{P} - \tilde{a}^{(\,0,1)\dag}M^{\mbsu{RPA}}_{\vphd}.
\label{defppt}
\ee
The absence of the last term of Eq. (\ref{defppt}) in the operator $\tilde{P}$
leads to the incomplete elimination of the ``ghost'' part of
$R^{\mbsu{RPA}}_{\vphd}(\omega)$ in this method.
Indeed, from Eqs. (\ref{rfsb}), (\ref{a01e1}), (\ref{a02e1}), (\ref{deftproj}),
and (\ref{ta01e1}) it follows that
\be
\tilde{P} R^{\mbsu{RPA(spur.)}}_{\vphd}(\omega) =
-\frac{\tilde{a}^{(\,0,1)\dag}}{\omega}\,.
\label{ptrfs}
\ee
The remainder determined by the r.h.s. of Eq. (\ref{ptrfs})
does not contribute into the strength function defined by Eqs. (\ref{sfdef}) and
(\ref{poldef}) in the case of the local external-field operators $Q$ considered
in \cite{Mizuyama_2012}. However, the contribution of this remainder can be
non-zero in the other cases arising in the beyond mean-field models.

\section{Numerical results}
\label{sec:res}

In this section the results of the fully self-consistent calculations
of the electric dipole ($E1$) excitations performed in the models based on the
Skyrme EDF are presented.
Two main models are considered: RPA and the time blocking approximation (TBA)
which is the model of the ERPA type including $1p1h\otimes$phonon configurations
on top of the $1p1h$ configuration space of the RPA. The general scheme of
the TBA is described by Eqs. (\ref{rferpa1}), (\ref{def:bw}), and (\ref{def:w})
of Sec. \ref{sec:erpa}. The detailed formulas of the self-consistent version of this model
are given in Refs. \cite{Lyutorovich_2016,Tselyaev_2016}.
Numerical details of the calculations are the following. The RPA and TBA equations
were solved in the representation of the discrete {\it p-h} basis
obtained from the solution of the Skyrme-Hartree-Fock equations
(\ref{speqs}) and (\ref{frpa}) with the box boundary condition.
The basis included all the hole states and all the particle states
of the single-particle spectrum with the energies
$\ve^{\vphu}_p < \ve_{\mbss{max}}$.
In what follows two versions of the RPA and TBA are used. In the discrete versions
(DRPA and DTBA) the effect of the single-particle continuum is not included.
In the continuum versions (CRPA and CTBA) this effect is included in the
discrete basis representation according to the method described in \cite{Tselyaev_2016}.

\subsection{SDM results in the RPA}
\label{sec:sdmrpa}

\begin{figure}[]
\begin{center}
\includegraphics*[trim=2cm 0cm 0cm 2cm,clip=true,scale=0.38,angle=90]{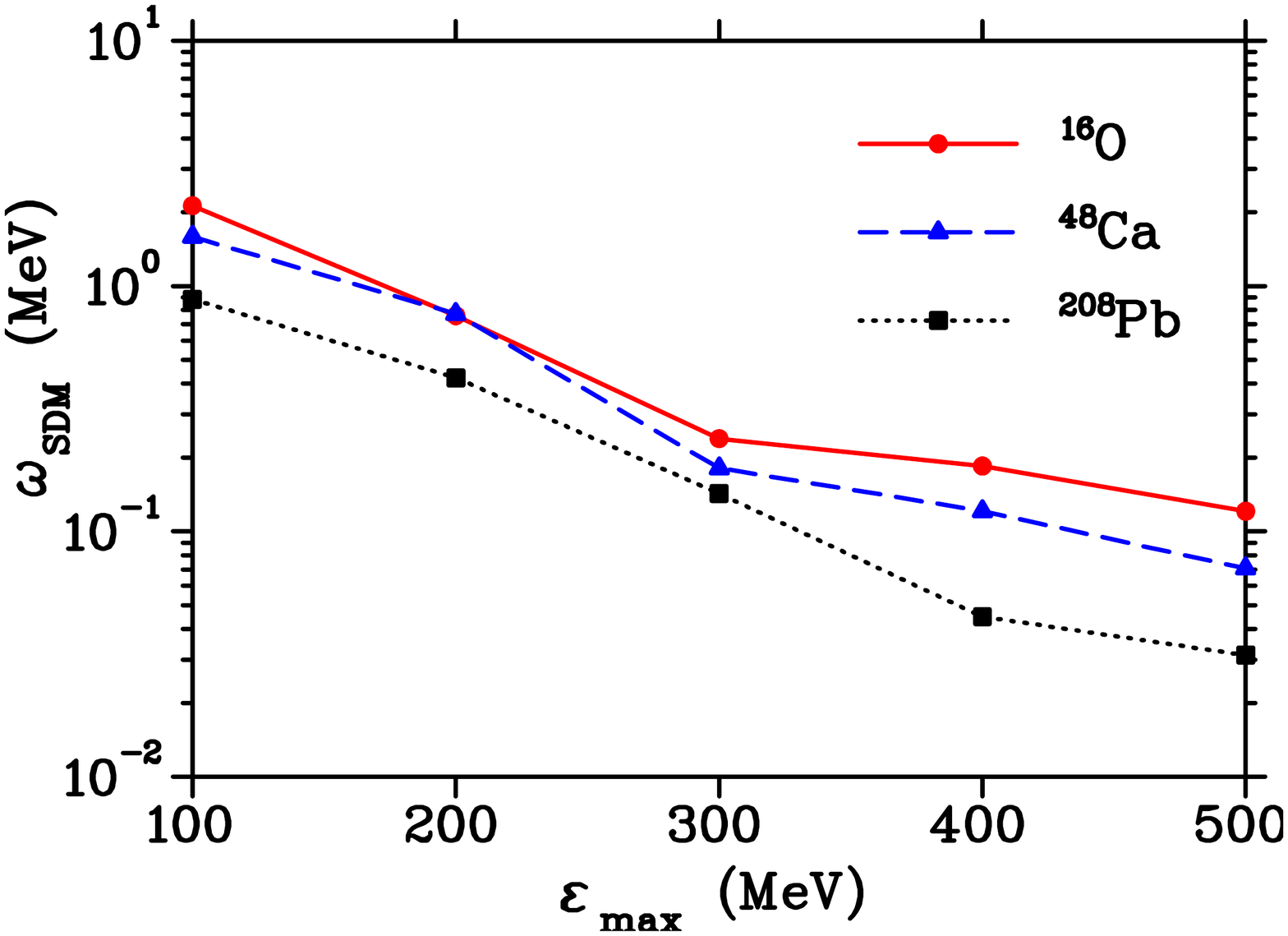}
\end{center}
\caption{\label{fig:esdm}
Dependence of the energy of the spurious dipole mode $\omega_{\mbss{SDM}}$
on the maximum energy of the states of the single-particle spectrum $\ve_{\mbss{max}}$.
Calculations within the fully self-consistent DRPA based on the
Skyrme-EDF parametrization SLy4 for $^{16}$O, $^{48}$Ca, and $^{208}$Pb.
}
\end{figure}

In the calculations of the $E1$ excitations within the self-consistent
RPA and the RPA-based models, the deviation of the SDM energy ($\omega_{\mbss{SDM}}$)
from zero can serve as a criterion of accuracy of the calculation scheme.
The rigorous equality $\omega_{\mbss{SDM}}=0$ implies the full self consistency,
that is exact fulfillment of Eqs. (\ref{speqs}) and (\ref{frpa}),
however the value of $\omega_{\mbss{SDM}}$ strongly depends also on the numerical
details, in particular on the size of the {\it p-h} basis determined by the parameter
$\ve_{\mbss{max}}$.
In Fig. \ref{fig:esdm}, the dependence of $\omega_{\mbss{SDM}}$ on the value of
$\ve_{\mbss{max}}$ is shown. The results were obtained in the DRPA for three
doubly magic nuclei: $^{16}$O, $^{48}$Ca, and $^{208}$Pb.
The Skyrme-EDF parametrization SLy4 \cite{Chabanat_1998} was used.
The box radius was taken to be 15~fm for $^{16}$O and $^{48}$Ca and 18~fm for $^{208}$Pb.
At the increase of $\ve_{\mbss{max}}$ from 100 to 500 MeV,
the value of $\omega_{\mbss{SDM}}$ decreases
from 2.13 to 0.12 MeV in $^{16}$O,
from 1.59 to 0.07 MeV in $^{48}$Ca, and
from 0.88 to 0.03 MeV in $^{208}$Pb.
The decrease of $\omega_{\mbss{SDM}}$ is monotonic and almost exponential
at $\ve_{\mbss{max}} <$ 300 MeV, but is slowed down at greater $\ve_{\mbss{max}}$.
In particular, $\omega_{\mbss{SDM}}\approx$ 3 keV in $^{16}$O at
$\ve_{\mbss{max}} =$ 2000 MeV.

The separation of the SDM from the physical modes can be estimated with the
help of the ratio of the reduced probability of the $E1$ transition for the SDM,
$B(E1)_{\mbss{SDM}}$, to the reduced probability $B(E1)_{\mbss{max}}$ for the DRPA state
having the largest $B(E1)$ in the given strength distribution.
For the effective $E1$ operator defined by Eqs. (\ref{qe1eff}) and (\ref{feffiv}),
the limiting value of $B(E1)_{\mbss{SDM}}$ is equal to zero.
The calculated values of $B(E1)_{\mbss{SDM}}/B(E1)_{\mbss{max}} \lesssim$
1.5$\times$10$^{-5}$ in the case of $\ve_{\mbss{max}} =$ 100 MeV
and $\lesssim$ 3.3$\times$10$^{-9}$ in the case of $\ve_{\mbss{max}} =$ 500 MeV
for all three nuclei under consideration.
Thus, a fairly good separation of the SDM is achieved in the DRPA already at
$\ve_{\mbss{max}} =$ 100 MeV.

\subsection{Center-of-mass motion in the response function formalism}
\label{sec:cmc}

In the ERPA theories, separation of the SDM is attained with the help of the method
described in Sec. \ref{sec:erpa}. Consider its implementation in the TBA.
The efficiency of this method can be estimated by comparing the calculated response
of the nucleus to the external field represented by the operator of the
center-of-mass coordinate (c.m.c.)
\be
\bfbq = \frac{m}{M^{\vphu}_0}\bfr
\label{qcmc}
\ee
with the known exact result.

From Eqs. (\ref{poldef}), (\ref{rf2t}), (\ref{rfphys}), (\ref{rfsb}),
(\ref{zszer})--(\ref{a02mr}) it follows that the polarizability
$\Pi (\omega)$ corresponding to the operator (\ref{qcmc}) in the self-consistent
RPA is equal to
\be
\Pi_{\,\mbss{c.m.}}(\omega) = \frac{3\hbar^2}{M^{\vphu}_0 \omega^2}.
\label{polcmc}
\ee
As follows from Eqs. (\ref{rfbperpa}), (\ref{rfberpa1}), and (\ref{rfberpa2}),
the same result for the polarizability is obtained in the self-consistent ERPA+
described in Sec. \ref{sec:erpa}, that is in the case of the response function
$R(\omega)={R}^{\mbsu{ERPA+}}_{\vphd}(\omega)$ in Eq. (\ref{poldef}).
On the other hand, in the ERPA without projection the c.m.c. polarizability does not
coincide with Eq. (\ref{polcmc}) in the general case.
However, the result expressed by Eq. (\ref{polcmc}) is not exactly achieved
even in the self-consistent RPA and ERPA+
calculations because of the numerical limitations and inaccuracies.
The measure of deviation from this exact expression can be determined
with the help of the function
\be
D_{\mbss{c.m.}}(E) = |\,\Pi(E + i\Delta)/\Pi_{\,\mbss{c.m.}}(E + i\Delta) - 1\,|\,,
\label{def:dcm}
\ee
where $\Pi(\omega)$ is the calculated polarizability, $E$ is the real energy
variable, and $\Delta$ is the smearing parameter.

\begin{figure}[]
\begin{center}
\includegraphics*[trim=2cm 0cm 0cm 2cm,clip=true,scale=0.38,angle=90]{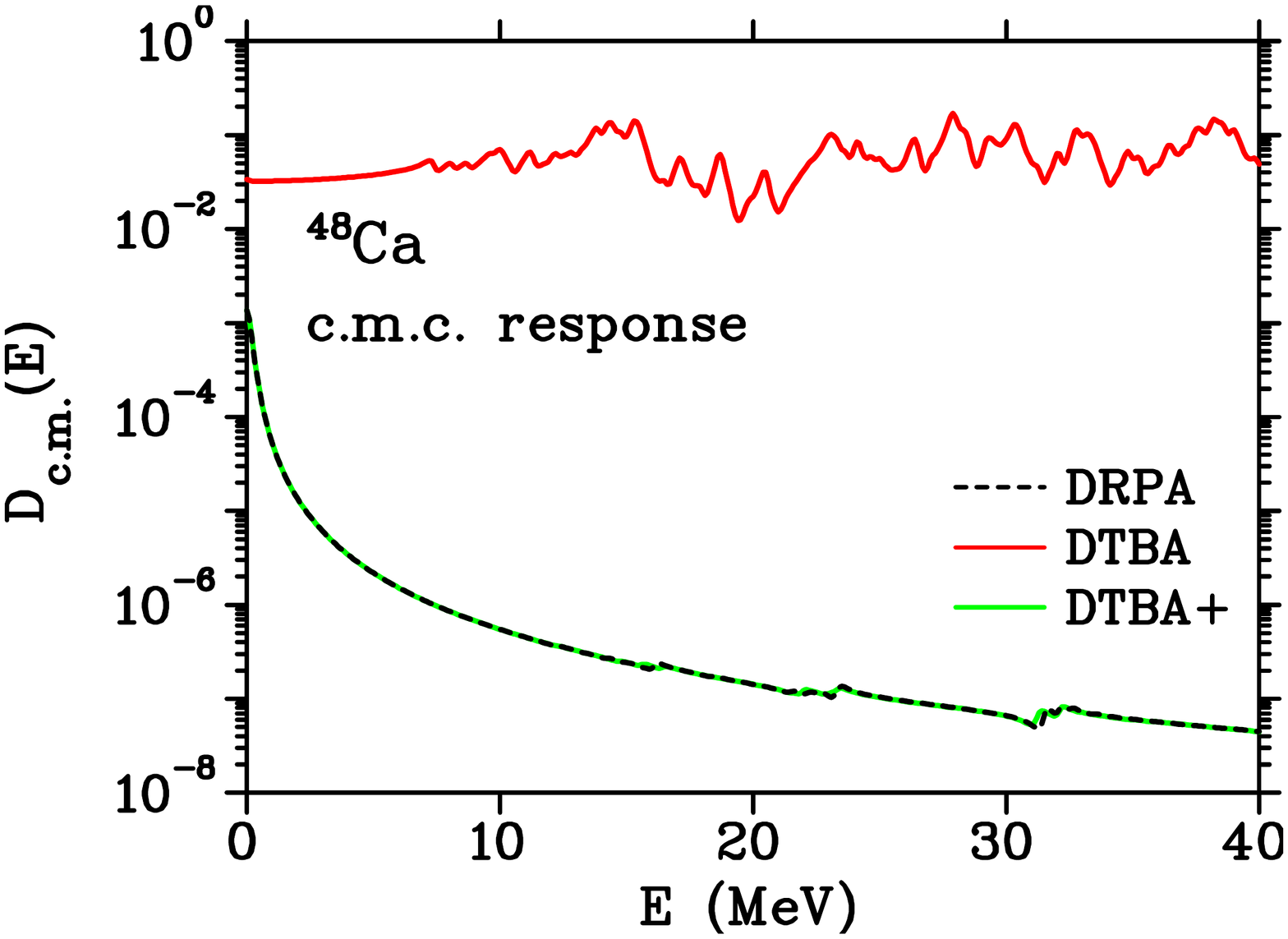}
\end{center}
\caption{\label{fig:e1cm}
Energy dependence of the function $D_{\mbss{c.m.}}(E)$, Eq. (\ref{def:dcm}),
as the measure of deviation of the calculated polarizability from the exact
formula (\ref{polcmc}) for the response of the nucleus to the external field
represented by the operator of the center-of-mass coordinate (c.m.c.).
Calculations for $^{48}$Ca within the fully self-consistent DRPA (black dashed line),
DTBA (red full line), and DTBA+ (green full line) based on the
Skyrme-EDF parametrization SLy4.
The smearing parameter $\Delta =$ 200 keV was used.
}
\end{figure}

In Fig.~\ref{fig:e1cm}, the calculated dependencies $D_{\mbss{c.m.}}(E)$
for the nucleus $^{48}$Ca
are shown for three fully self-consistent models: DRPA (black dashed line),
DTBA (red full line), and DTBA+ (green full line) based on the Skyrme-EDF
parametrization SLy4.
The {\it p-h} basis included all single-particle states up to the energy
$\ve_{\mbss{max}} =$ 1000 MeV that gives $\omega_{\mbss{SDM}} =$ 7 keV.
The phonon basis in the DTBA and DTBA+ included three low-lying most collective
RPA states with $L^{\pi}$ = $2^+$, $3^-$, and $5^-$, having the energies equal to
3.37, 5.61, and 6.03 MeV, respectively. In the calculations of $D_{\mbss{c.m.}}(E)$,
the smearing parameter $\Delta =$ 200 keV was used.
The DRPA and DTBA+ results practically coincide with each other that demonstrates
well performance of the projection method used in the DTBA+.
The maximum value of $D_{\mbss{c.m.}}(E)$ in both these models is attained
at $E=0$ and is about $10^{-3}\approx \omega^2_{\mbsu{SDM}}/\Delta^2$.
At the large values of $E$, the function
$D_{\mbss{c.m.}}(E)\approx \omega^2_{\mbsu{SDM}}/E^2$
and becomes less than $10^{-6}$ at $E>10$ MeV.
This dependence of $D_{\mbss{c.m.}}(E)$ is well approximated by the function
\be
\tilde{D}_{\mbss{c.m.}}(E) = \frac{\omega^2_{\mbsu{SDM}}}
{\sqrt{(E^2 - \omega^2_{\mbsu{SDM}} - \Delta^2)^2 + 4 E^2 \Delta^2}},
\label{def:dcma}
\ee
which is obtained from the definitions (\ref{qcmc})--(\ref{def:dcm}) and
from Eqs. (\ref{poldef}), (\ref{rfphys}), and (\ref{zssdef})
at finite $\omega^{\vphu}_0 = \omega_{\mbss{SDM}}$.
Thus, the degree of separation of the SDM in the ERPA+ calculations
is determined by the proximity of $\omega_{\mbss{SDM}}$ to zero.

In the DTBA without projection [but with subtraction according to Eq. (\ref{def:bw})],
the values of $D_{\mbss{c.m.}}(E)$ oscillate within the range from 0.01 up to 0.17
in the considered energy interval that shows the existing residual coupling of the SDM
with the physical modes in this model.

\subsection{Elimination of the SDM in realistic TBA calculations}
\label{sec:realc}

Consider application of the method of the elimination of the SDM
in the realistic calculations of $E1$ excitations in $^{48}$Ca, $^{48}$Ni, and $^{208}$Pb
within the renormalized version of the TBA (RenTBA) developed in Ref. \cite{Tselyaev_2018}.
The final equations of both the RenTBA and the TBA have the form of Eqs. (\ref{rferpa1}),
(\ref{def:bw}), and (\ref{def:w}) of Sec. \ref{sec:erpa}, but the phonon space in the RenTBA
is determined by the system of nonlinear equations. The details of its solution
are described in \cite{Tselyaev_2018} and here are the same as in Ref. \cite{Tselyaev_2020}
except for the space of the phonon renormalization which in the
present paper is common for the phonons of the electric and magnetic types.

\subsubsection{$E1$ excitations in $^{48}$Ca and $^{48}$Ni}
\label{sec:realc-1}

In the calculations of giant resonances in the light nuclei and especially in the light
exotic nuclei, the contribution of the single-particle continuum is large.
For this reason, it was included in the calculations for $^{48}$Ca and proton-rich $^{48}$Ni.
The abbreviations CTBA and CTBA+ refer here to the RenTBA results.
The details of the calculation scheme are the following.
The Skyrme-EDF parametrization SV-m64k6 was used. This parametrization was suggested
in Ref. \cite{Lyutorovich_2012} (where it was denoted as SV-m64-O)
for the description of the giant dipole resonances (GDR) in the light nuclei.
To avoid the spin instability of the ground state, the so-called spin surface terms
of the EDF have been omitted (see Ref. \cite{Tselyaev_2019} for more details).
It corresponds to the option $\eta^{\vphu}_{\Delta s} = 0$ in terms of
Ref. \cite{Tselyaev_2019}.
The {\it p-h} basis was restricted by the parameter $\ve_{\mbss{max}} =$ 100 MeV.
The box radius was equal to 15 fm in $^{48}$Ca and 18 fm in $^{48}$Ni.
It gives $\omega_{\mbss{SDM}} =$ 1.58 MeV in $^{48}$Ca and 1.53 MeV in $^{48}$Ni.
%
The resulting phonon space of the RenTBA in $^{48}$Ca
included 8 phonons of the electric type with multipolarities $2 \leqslant L \leqslant 6$
and 8 phonons of the magnetic type with multipolarities $1 \leqslant L \leqslant 5$.
The phonon space in $^{48}$Ni included
14 phonons of the electric type with multipolarities $2 \leqslant L \leqslant 6$
and 7 phonons of the magnetic type with multipolarities $2 \leqslant L \leqslant 5$.
All the obtained phonon's energies are less than 12 MeV.

\begin{figure}[h!]
\begin{center}
\includegraphics*[trim=2cm 0cm 0cm 2cm,clip=true,scale=0.38,angle=90]{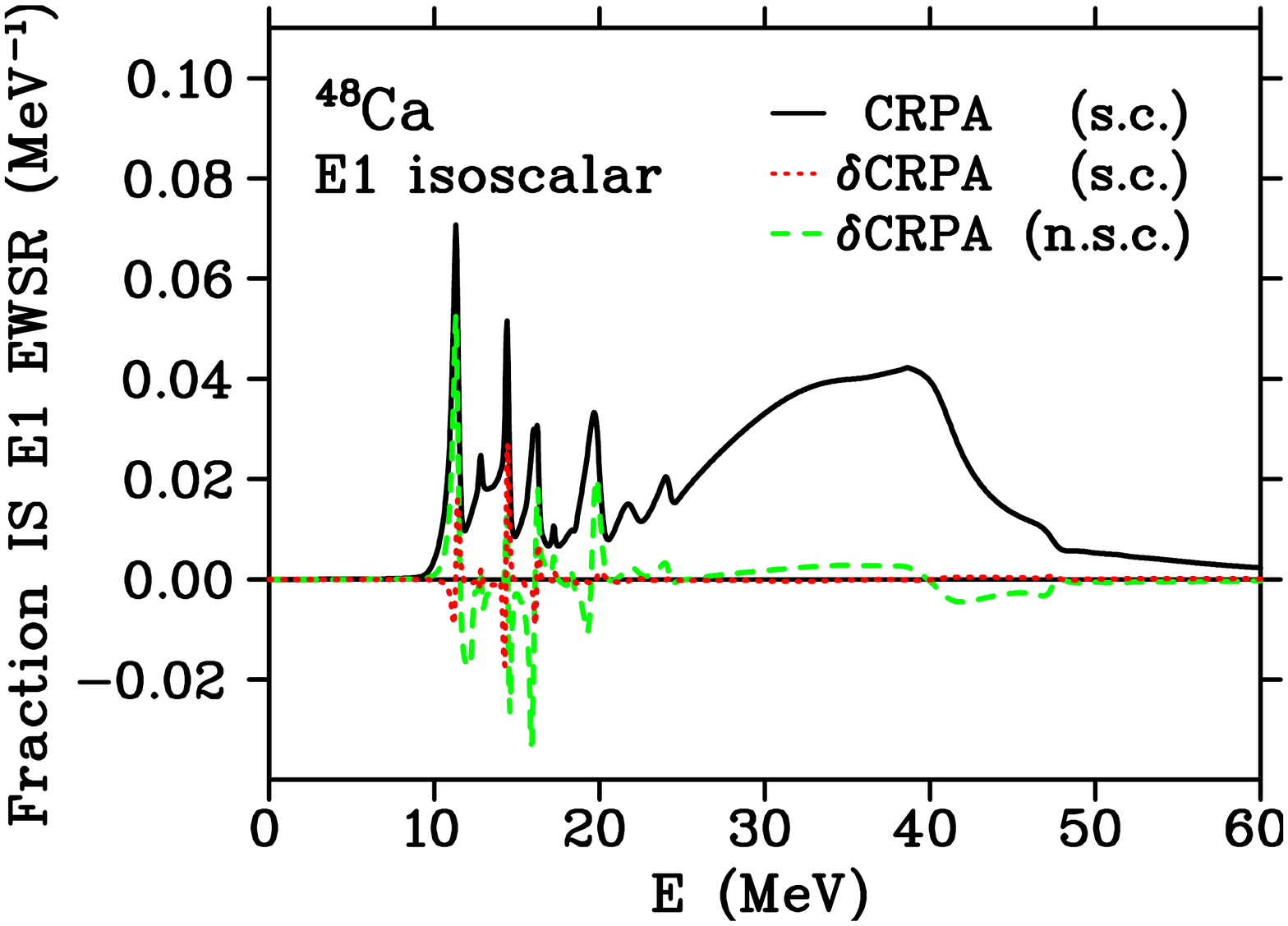}
\end{center}
\caption{\label{fig:e1isrpa}
Strength distribution of the IS $E1$ resonance in $^{48}$Ca calculated
within the CRPA based on the Skyrme-EDF parametrization SV-m64k6.
The black full line corresponds to the fully self-consistent CRPA with
$\ve_{\mbss{max}} =$ 100 MeV.
The red dotted line represents the difference between the results obtained
in fully self-consistent CRPA with $\ve_{\mbss{max}} =$ 100 MeV and
$\ve_{\mbss{max}} =$ 1000 MeV.
The green dashed line represents the difference between the results obtained
in fully self-consistent and non-self-consistent CRPA with $\ve_{\mbss{max}} =$ 100 MeV.
The smearing parameter $\Delta =$ 100 keV was used.
See text for more details.}
\end{figure}

In general, the numerical scheme described above is conventional for most
TBA and RenTBA calculations.
Nevertheless, consider the influence of some elements of this scheme on the results
in the case of the isoscalar (IS) $E1$ excitations.
Dependence of the SDM energy on the parameter $\ve_{\mbss{max}}$ was analyzed in
Sec.~\ref{sec:sdmrpa}.
Dependence of the strength distributions of the giant IS $E1$ resonances
on this parameter is much less pronounced.
Figure \ref{fig:e1isrpa} shows such a distribution in $^{48}$Ca calculated
in the CRPA in terms of the fraction of the IS $E1$ energy-weighted sum rule
(IS $E1$ EWSR) determined by the function
\be
F(E) = E\,S(E)/m^{\mbsu{RPA}}_1\,,
\label{def:frewsr}
\ee
where $S(E)$ is the strength function (\ref{sfdef}),
\be
m^{\mbsu{RPA}}_1 = \int_0^{\infty} E\,S^{\mbsu{RPA}}(E)\,dE
\label{def:ewsr}
\ee
is the first moment of the RPA strength function $S^{\mbsu{RPA}}(E)$.

The function $S(E)$ was calculated for the effective IS $E1$ operator determined by
Eqs. (\ref{qe1eff}) and (\ref{feffis}). For this operator, the analytic formula
for $m^{\mbsu{RPA}}_1$ is known (RPA EWSR, see Ref. \cite{Giai_1981}).
The black full line in Fig.~\ref{fig:e1isrpa}
represents the function $F^{\,\mbsu{s.c.}}_{100}(E)$
calculated according to Eq. (\ref{def:frewsr})
within the fully self-consistent CRPA with $\ve_{\mbss{max}} =$ 100 MeV
and the smearing parameter $\Delta =$ 100 keV.
The analogous function $F^{\,\mbsu{s.c.}}_{1000}(E)$ was calculated with
$\ve_{\mbss{max}} =$ 1000 MeV
(note that $\omega_{\mbss{SDM}} =$ 11 keV for this value of $\ve_{\mbss{max}}$).
The curves corresponding to $F^{\,\mbsu{s.c.}}_{100}(E)$ and $F^{\,\mbsu{s.c.}}_{1000}(E)$
are hardly distinguishable from each other in the given scale, so only the difference
\be
\delta F^{\,\mbsu{s.c.}}(E) = F^{\,\mbsu{s.c.}}_{100}(E) - F^{\,\mbsu{s.c.}}_{1000}(E),
\ee
denoted as $\delta$CRPA (s.c.), is shown in Fig.~\ref{fig:e1isrpa} by the red dotted line.
This difference is not negligible only for three narrow peaks in the region of
11--16.2 MeV which are shifted down in the function $F^{\,\mbsu{s.c.}}_{1000}(E)$
by the value $\delta E$ from 30 to 130 keV.
The difference $\delta F^{\,\mbsu{s.c.}}(E)$ practically vanishes in the region
of the giant IS $E1$ resonance at $E >$ 20 MeV.

For comparison, the function $F^{\,\mbsu{n.s.c.}}_{100}(E)$ for the IS $E1$ excitations
in $^{48}$Ca was calculated by the formula (\ref{def:frewsr})
within non-self-consistent CRPA with $\ve_{\mbss{max}} =$ 100 MeV.
In this calculation the spin-orbit and Coulomb contributions into the residual interaction
$V$ in Eq. (\ref{orpa}) have been omitted
that leads to the increase of $\omega_{\mbss{SDM}}$ from 1.58 to 2.74 MeV.
The difference
\be
\delta F^{\,\mbsu{n.s.c.}}(E) = F^{\,\mbsu{s.c.}}_{100}(E) - F^{\,\mbsu{n.s.c.}}_{100}(E)
\ee
is shown in Fig. \ref{fig:e1isrpa} by the green dashed line denoted as
$\delta$CRPA (n.s.c.).
As can be seen, this difference does not vanish in the wide region from 10 to 50 MeV.

\begin{figure}[h!]
\begin{center}
\includegraphics*[trim=1.8cm 5cm 0cm 0cm,clip=true,scale=0.5,angle=0]{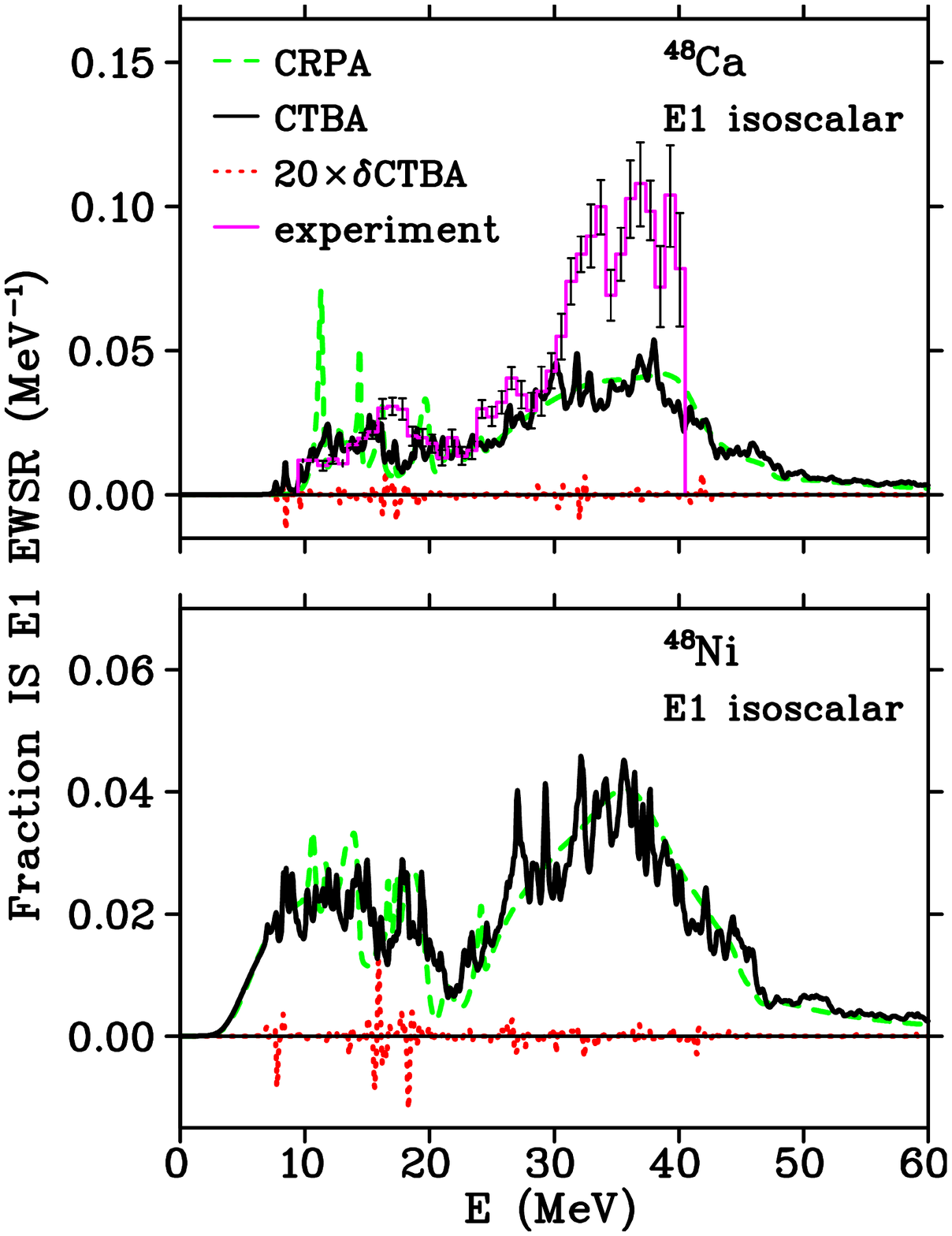}
\end{center}
\caption{\label{fig:e1is}
Upper panel:
strength distributions of the isoscalar $E1$ excitations in $^{48}$Ca calculated
within the CRPA (green dashed line) and CTBA (black full line)
based on the Skyrme-EDF parametrization SV-m64k6.
The red dotted line represents the difference $\delta$CTBA between the CTBA+ and CTBA
results multiplied by the factor of 20.
The smearing parameter $\Delta =$ 100 keV was used.
Experimental data (magenta full line with error bars) are taken from Ref. \cite{Lui_2011}.
Lower panel: same as in the upper panel but for $^{48}$Ni.
}
\end{figure}

The results of the CTBA calculations of the IS $E1$ excitations in $^{48}$Ca and
$^{48}$Ni are shown in Fig. \ref{fig:e1is}. They are presented in terms of the
fractions of the RPA EWSR determined by Eq. (\ref{def:frewsr}).
The strength functions in Eq. (\ref{def:frewsr}) were calculated with the same
smearing parameter $\Delta =$ 100 keV as in Fig. \ref{fig:e1isrpa}.
The narrow peaks of the CRPA strength distribution in $^{48}$Ca in the region below 20 MeV
are strongly fragmented in the CTBA in agreement with the experimental data for
this nucleus from Ref. \cite{Lui_2011}.
However, the theory does not describe the large increase of IS $E1$ strength above
30 MeV found experimentally. This discrepancy takes place also in the other
RPA calculations of the IS $E1$ giant resonance in $^{48}$Ca, see Ref. \cite{Anders_2013}.

The effect of the elimination of the SDM in the CTBA+ can be estimated with the help
of the difference (denoted as $\delta$CTBA in Fig. \ref{fig:e1is}) between the functions
$F(E)$ calculated within the CTBA+ and CTBA.
As can be seen from Fig.~\ref{fig:e1is}, the relative difference is very small
in both nuclei and only scaled $\delta$CTBA (multiplied by the factor of 20) is visible.
This difference is much less than the differences between the CRPA distributions shown
in Fig. \ref{fig:e1isrpa} and between the DTBA+ and DTBA results for the c.m.c.
response shown in Fig. \ref{fig:e1cm}. The latter fact needs an explanation.

\begin{figure}[h!]
\begin{center}
\includegraphics*[trim=2cm 0cm 0cm 2cm,clip=true,scale=0.38,angle=90]{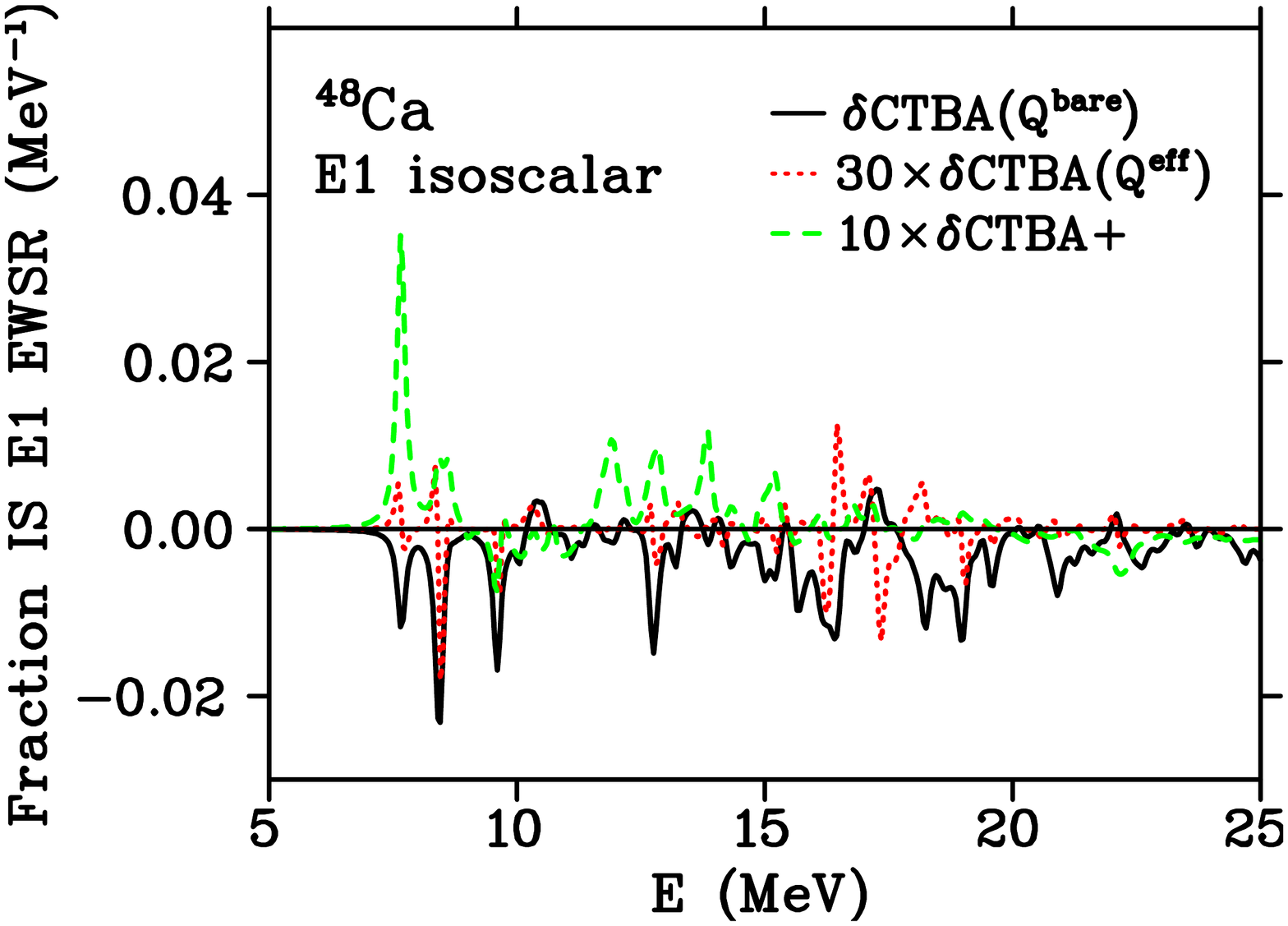}
\end{center}
\caption{\label{fig:qbe}
Differences of the
strength distributions of the IS $E1$ resonance in $^{48}$Ca calculated
within the CTBA based on the Skyrme-EDF parametrization SV-m64k6.
The black full line corresponds to the difference $\delta$CTBA between the CTBA+ and CTBA
results obtained for the bare IS $E1$ operator (\ref{qe1b}), (\ref{fe1isb}).
The red dotted line represents analogous difference for the effective IS $E1$ operator
(\ref{qe1eff}), (\ref{feffis}) multiplied by the factor of 30.
The green dashed line represents the difference $\delta$CTBA+ between the CTBA+ results
obtained for the bare and effective IS $E1$ operators multiplied by the factor of 10.
The smearing parameter $\Delta =$ 100 keV was used.
}
\end{figure}

It should be specified that the results shown in Fig. \ref{fig:e1is}
were obtained for the effective IS $E1$ operator determined by
Eqs. (\ref{qe1eff}) and (\ref{feffis}). But this operator itself suppresses the spurious
admixtures in the excited states as follows from its general definition (\ref{defqeff}).
For this reason, the effect of the projection method used in the CTBA+ is better seen
in the calculations of the IS $E1$ response for the bare operator determined by
Eqs. (\ref{qe1b}) and (\ref{fe1isb}).
However, in this case the main component of the SDM (fragmented in the TBA without projection)
gives the large background in the strength function
if the smearing parameter $\Delta$ in Eq. (\ref{sfdef}) is not very small.
The calculations for $^{48}$Ca were performed with $\Delta =$ 100 keV,
and to remove this background the contribution of the main component of the SDM
into the strength functions
was eliminated with the help of subtraction of the SDM term with the minimum energy
from the polarizability (\ref{poldef}).
The difference $\delta$CTBA$(\bfbq^{\mbsu{bare}})$ between the functions
$F(E)$ calculated within the CTBA+ and CTBA for the bare IS $E1$ operator
is shown in Fig. \ref{fig:qbe} by the black full line.
As can be seen, the absolute values of $\delta$CTBA$(\bfbq^{\mbsu{bare}})$
are on the average much larger than the values of the analogous difference
$\delta$CTBA$(\bfbq^{\mbsu{eff}})$ calculated for the effective IS $E1$ operator
and shown in Fig. \ref{fig:qbe} by the red dotted line.
On the other hand, the difference $\delta$CTBA+ between the CTBA+ results
obtained for the IS $E1$ operators $\bfbq^{\mbsu{bare}}$ and $\bfbq^{\mbsu{eff}}$
(shown by the green dashed line) is small.

Differences of the strength distributions shown in Fig.~\ref{fig:qbe} can be
quantified with the help of the following relative mean-square deviation
\be
|| \delta S ||_{\mbss{R}} = || \delta S ||/|| S ||\,,
\label{def:dlts}
\ee
with
\be
|| \delta S ||^2 = \int_{E_1}^{E_2} \bigl( \tilde{S}(E) - S(E) \bigr)^2 dE\,,
\label{def:ids2}
\ee
\be
|| S ||^2 = \int_{E_1}^{E_2} S^2(E)\,dE\,,
\label{def:is2}
\ee
where $S(E)$ is the strength function calculated within the CTBA+
(for the IS $E1$ operator $\bfbq^{\mbsu{eff}}$ in the case of $\delta$CTBA+)
and $\tilde{S}(E)$ is the strength function calculated within the versions of
the CTBA and CTBA+ shown in Fig. \ref{fig:qbe}.

For the energy intervals 5--25 and 0--60 MeV in
Eqs. (\ref{def:ids2}) and (\ref{def:is2}),
the values of $|| \delta S ||_{\mbss{R}}$ are listed in Table~\ref{tab:ds}.
The large values of $|| \delta S ||_{\mbss{R}}$ for the difference
$\delta$CTBA$(\bfbq^{\mbsu{bare}})$ show that the coupling of the SDM
with physical modes in the CTBA response function for the IS $E1$ excitations
is quite appreciable. However, the small values of $|| \delta S ||_{\mbss{R}}$ for
$\delta$CTBA$(\bfbq^{\mbsu{eff}})$ and $\delta$CTBA+ suggest, first, that
this coupling is effectively eliminated in the CTBA+ response function and,
second, that the strength of the SDM fragments resulting from this coupling in the
excitation spectrum of the CTBA is strongly suppressed by the effective
IS $E1$ operator $\bfbq^{\mbsu{eff}}$, as mentioned above.

\begin{table}
\caption{\label{tab:ds}
Integral characteristics of the
differences of the strength distributions shown in Fig. \ref{fig:qbe}.
The relative mean-square deviation $|| \delta S ||_{\mbss{R}}$ is determined by
Eqs. (\ref{def:dlts})--(\ref{def:is2}).
}
\begin{ruledtabular}
\begin{tabular}{cccc}
interval & & $|| \delta S ||_{\mbss{R}}$ & \\
(MeV) & $\delta$CTBA$(\bfbq^{\mbsu{bare}})$
& $\delta$CTBA$(\bfbq^{\mbsu{eff}}) \vphantom{\frac{A^{B^C}}{D}}$
& $\delta$CTBA+ \\
\hline
 5--25 & 0.43 & 0.008 & 0.05 \\
 0--60 & 0.31 & 0.006 & 0.04 \\
\end{tabular}
\end{ruledtabular}
\end{table}

\begin{figure}[]
\begin{center}
\includegraphics*[trim=1.8cm 5cm 0cm 0cm,clip=true,scale=0.5,angle=0]{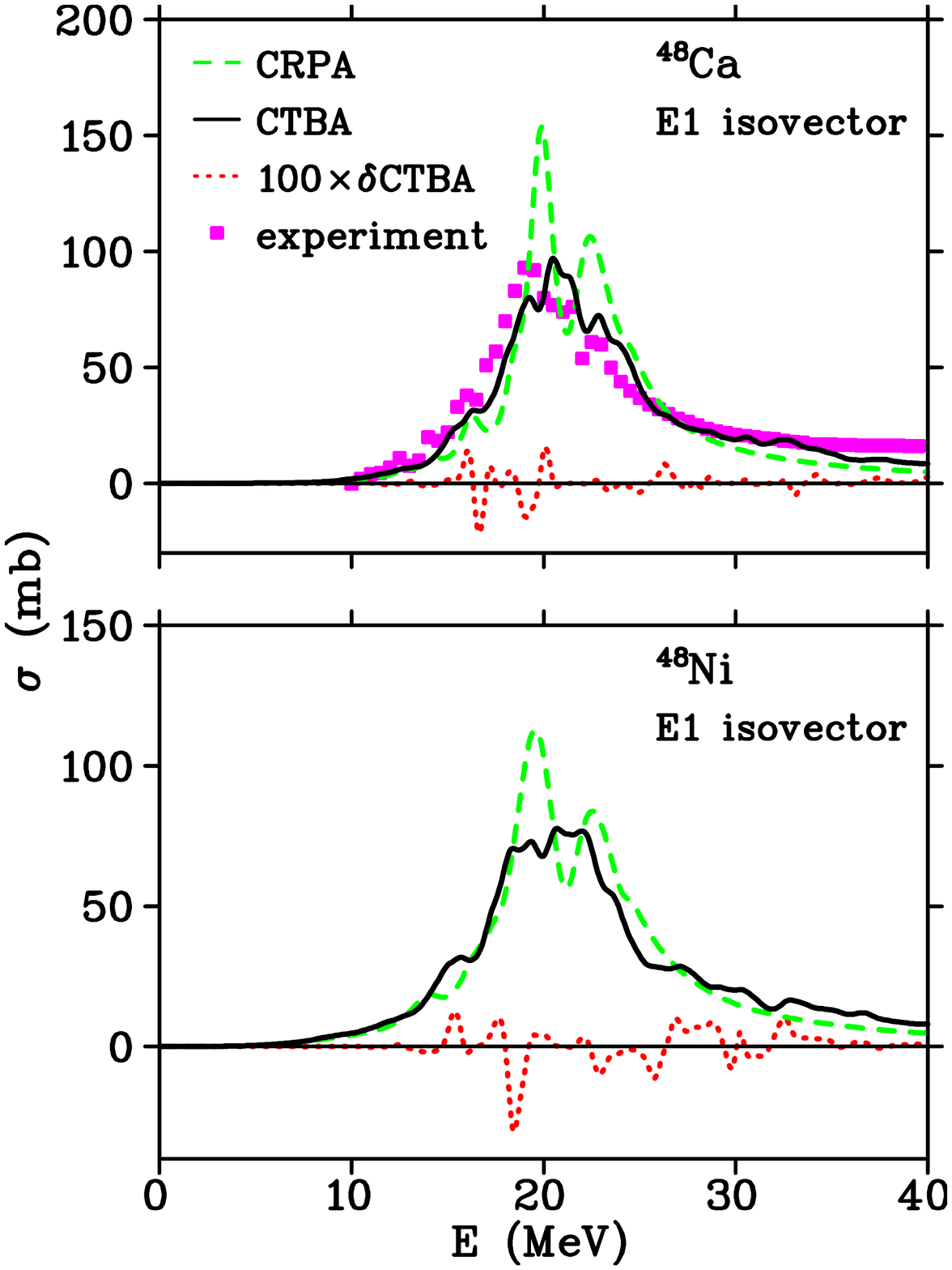}
\end{center}
\caption{\label{fig:e1iv}
Upper panel:
the $E1$ photoabsorption cross section for $^{48}$Ca calculated within
the CRPA (green dashed line) and CTBA (black full line)
based on the Skyrme-EDF parametrization SV-m64k6.
The red dotted line represents the difference $\delta$CTBA between the CTBA+ and CTBA
results multiplied by the factor of 100.
The smearing parameter $\Delta =$ 500 keV was used in the calculations of the
strength functions.
Experimental data from Ref. \cite{Erokhova_2003b} are represented by magenta squares.
Lower panel: same as in the upper panel but for $^{48}$Ni.
}
\end{figure}

In Fig.~\ref{fig:e1iv}, the results for the GDR in $^{48}$Ca and $^{48}$Ni are shown.
The $E1$ photoabsorption cross section shown in this figure
is connected with the strength function (\ref{sfdef}) for the effective
isovector $E1$ operator, Eqs. (\ref{qe1eff}) and (\ref{feffiv}),
by means of the known formula (see, e.g., \cite{Litvinova_2007}).
The strength functions for this operator were calculated with the smearing parameter
$\Delta =$ 500 keV.
The curve denoted as $100\times\delta$CTBA corresponds here to the difference
between the cross sections calculated within the CTBA+ and CTBA
multiplied by the factor of 100.
As in the case of the IS $E1$ excitations, this difference is very small.
It vanishes below 10 MeV where the isovector $E1$ strength in $^{48}$Ca
and $^{48}$Ni is absent.

The agreement between the theory and experiment in $^{48}$Ca is noticeably improved
in the CTBA as compared to the CRPA.
The form of the experimental $E1$ photoabsorption cross section is on the whole
reproduced in the CTBA, though the position of the main peak in the CTBA is
shifted upward by about 1.3 MeV with respect to the experiment.

\subsubsection{Pygmy dipole resonance in $^{208}$Pb}
\label{sec:realc-2}

Calculations of the pygmy dipole resonance (PDR) in $^{208}$Pb
were performed with the use of the Skyrme-EDF parametrization SKXm$_{-0.49}$
from Ref. \cite{Tselyaev_2020}. The spin-orbit and spin-spin parameters of
SKXm$_{-0.49}$ were fitted with the aim of description of the $M1$ resonance
in $^{208}$Pb within the RenTBA. The other parameters coincide with parameters
of the original Skyrme interaction SKXm \cite{Brown_1998}.
In the present paper, as mentioned above, the modified version of the renormalization
scheme of the RenTBA is used. In this case the results of Ref. \cite{Tselyaev_2020} for
the $M1$ resonance in $^{208}$Pb are reproduced at the slight change of the
spin-spin Landau-Migdal parameters $g$ and $g'$.
Note that these parameters do not affect the ground-state properties of the even-even
spherical nuclei.
In the calculations presented below the values $g=$ 0.086 and $g'=$ 0.87 are used
(the other parameters of SKXm$_{-0.49}$ are the same as in Ref. \cite{Tselyaev_2020}).
The box radius in $^{208}$Pb was equal to 18 fm, the parameter $\ve_{\mbss{max}} =$ 100 MeV.
It gives $\omega_{\mbss{SDM}} =$ 0.78 MeV.
The resulting phonon space of the RenTBA
included 124 phonons of the electric type with multipolarities $1 \leqslant L \leqslant 12$
and 73 phonons of the magnetic type with multipolarities $0 \leqslant L \leqslant 14$.
All the obtained phonon's energies are less than 11 MeV.

\begin{figure}[]
\begin{center}
\includegraphics*[trim=1.8cm 5cm 0cm 0cm,clip=true,scale=0.5,angle=0]{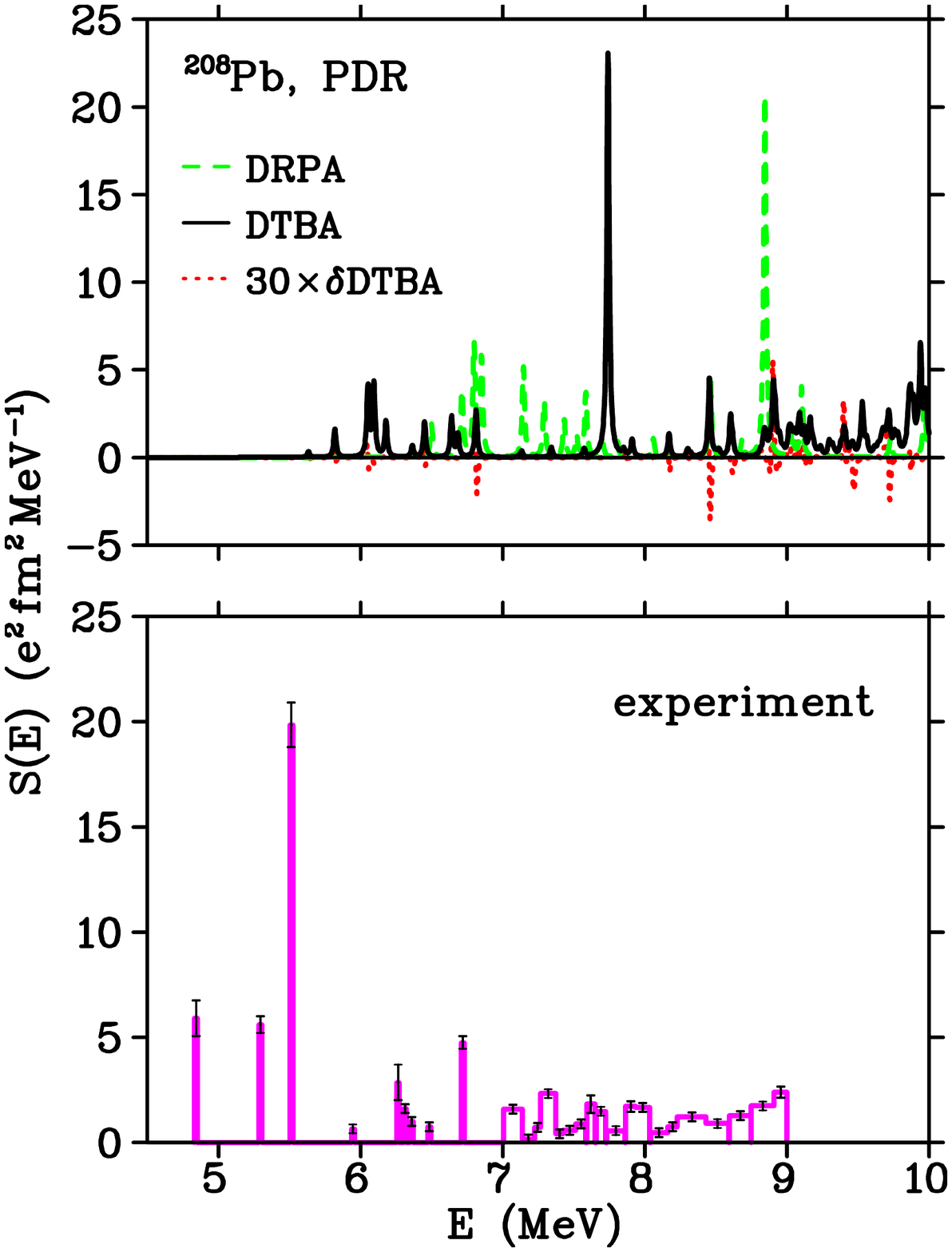}
\end{center}
\caption{\label{fig:pdr}
Upper panel:
pygmy dipole resonance in $^{208}$Pb calculated within
the DRPA (green dashed line) and DTBA (black full line)
based on the Skyrme-EDF parametrization SKXm$_{-0.49}$.
The red dotted line represents the difference $\delta$DTBA between the DTBA+ and DTBA
results multiplied by the factor of 30.
The smearing parameter $\Delta =$ 10 keV was used.
Lower panel:
experimental $E1$ strength distribution in $^{208}$Pb from Ref. \cite{Poltoratska12}.
See text for more details.}
\end{figure}

On the upper panel of Fig.~\ref{fig:pdr},
the strength distributions of the PDR in $^{208}$Pb
calculated within the fully self-consistent DRPA and DTBA are shown.
The single-particle continuum was not included because it plays a minor role here.
The strength functions for the effective isovector $E1$ operator determined by
Eqs. (\ref{qe1eff}) and (\ref{feffiv}) have been calculated with the small
smearing parameter $\Delta =$ 10 keV to expose the fine structure of the PDR.
The curve denoted as $30\times\delta$DTBA corresponds to the difference
between the functions $S(E)$ calculated within the DTBA+ and DTBA
multiplied by the factor of 30. This difference is small as in the cases of calculations
for $^{48}$Ca and $^{48}$Ni discussed above.

The experimental data from Ref. \cite{Poltoratska12} are presented
on the lower panel of Fig.~\ref{fig:pdr}
in the form of the distribution of the summed $B(E1)$ strengths in the energy bins divided
by the width of the bin $\Delta E$. Such a distribution corresponds to the strength function
$S(E)$ with the energy-dependent smearing parameter $\Delta$.
For the isolated $1^-$ states below 7 MeV, the value of $\Delta E$ was taken to be 20 keV
in accordance with the value of the smearing parameter in the DRPA and DTBA calculations.

\begin{figure}[]
\begin{center}
\includegraphics*[trim=2cm 0cm 0cm 2cm,clip=true,scale=0.38,angle=90]{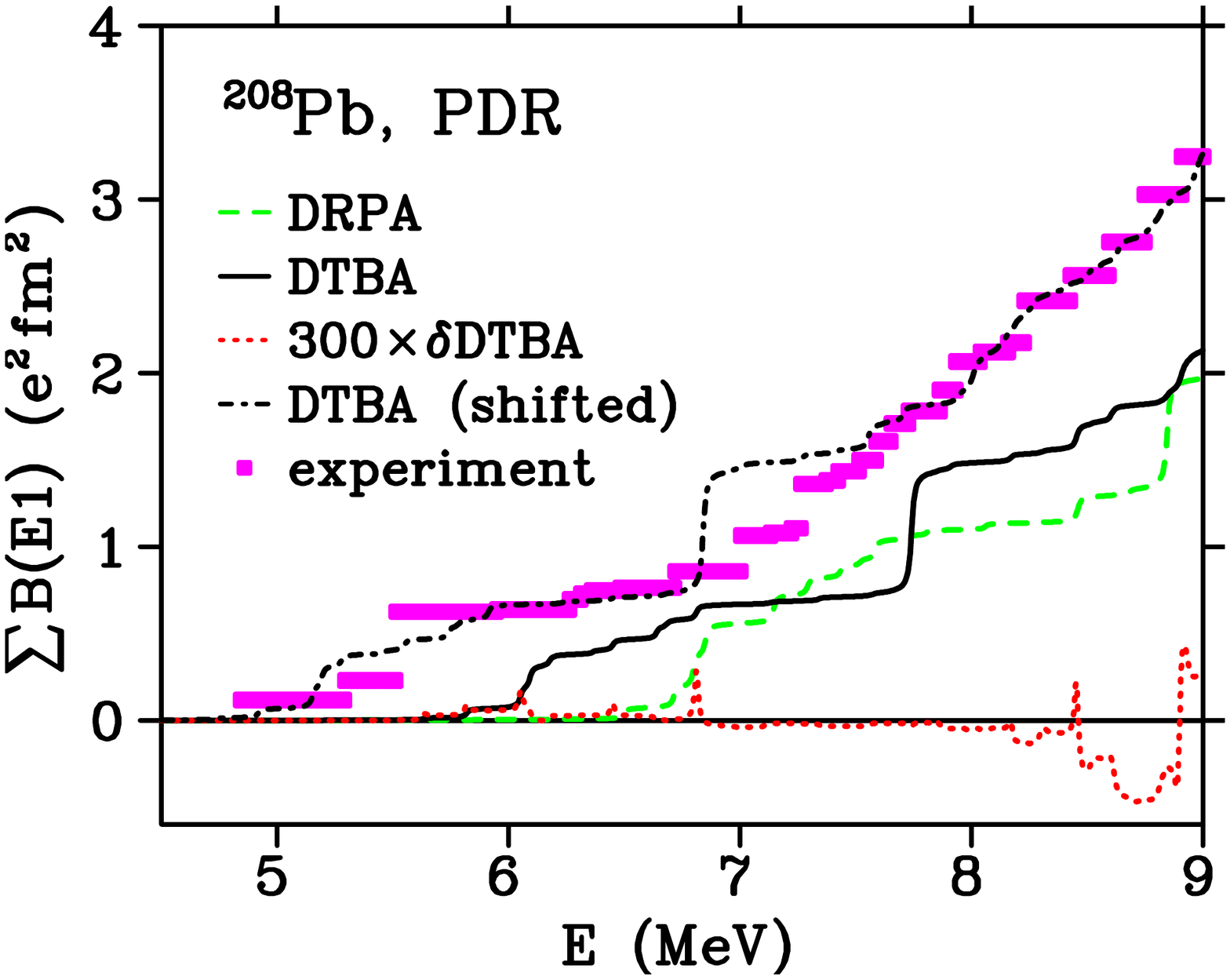}
\end{center}
\caption{\label{fig:sbpdr}
Summed strength of the $E1$ excitations in $^{208}$Pb in the region of the
pygmy dipole resonance calculated within
the DRPA (green dashed line) and DTBA (black full line)
based on the Skyrme-EDF parametrization SKXm$_{-0.49}$.
The red dotted line represents the difference $\delta$DTBA between the DTBA+ and DTBA
results multiplied by the factor of 300.
The black dash-dotted line represents the DTBA result shifted down by 0.9 MeV.
The experimental data from Ref. \cite{Poltoratska12} are represented by magenta squares.
}
\end{figure}

As was noted in Ref. \cite{Lyutorovich_2018}, the PDR in $^{208}$Pb can be divided into
two parts: the lower one from 4.8 to about 5.7 MeV, and the upper one
from 5.7 to 8.23 MeV. At higher energies, according to analysis of Ref. \cite{Poltoratska12},
the distribution of the $E1$ strength corresponds to the low-energy tail of the GDR.
However, the $E1$ strength below 5.7 MeV is absent both in the DRPA and in the DTBA
distributions shown in Fig.~\ref{fig:pdr}.
In general, it is a problem of many self-consistent calculations
of the PDR in $^{208}$Pb, see \cite{Lyutorovich_2018} for a more detailed discussion.
Though the comparison of the theory with the experimental data is not the aim
of the present paper, consider how the situation looks for the summed $E1$ strength
in $^{208}$Pb in the PDR region. The respective results are shown in Fig. \ref{fig:sbpdr}.
First, one can see that the difference between the DTBA+ and DTBA results for
the summed $E1$ strength is practically absent below 8 MeV and is negligibly small
in the interval 8--9 MeV. Second, the DTBA curve fairly well reproduces the data
(except for the region around 7 MeV) {\it if} this curve is shifted down by 0.9 MeV.
The DTBA itself gives the downward shift of the $E1$ strength as compared to the DRPA
but this shift is insufficient. One of the possible ways to diminish this discrepancy
between the theory and experiment is generalization of the RenTBA including so-called
ground-state correlations beyond the RPA
(see Refs. \cite{Kamerdzhiev_1997e,Kamerdzhiev_2004} for more details).

\section{Conclusions}
\label{sec:Conc}

In this work, the range of the problems related to the existence of
the spurious dipole mode (SDM) in the self-consistent nuclear-structure models
is considered.

An explicit form of the SDM terms of the RPA response function
was derived from the symmetry properties of the underlying energy-density functional.
This form was used to construct the projection operator $P$
which enables one
(i) to eliminate the SDM contributions from the RPA response function and
(ii) to eliminate coupling of the SDM with the physical modes in the
extended RPA (ERPA) theories.
The equation for the response function in the ERPA with projection
(ERPA+ in which the above-mentioned coupling is eliminated) is formulated.
It is shown that the action of the operator $P$ on the bare external-field
$E1$ operators yields the well-known isoscalar and isovector effective $E1$ operators.

Numerical examples of the application of the projection method are considered.
As the model of the ERPA type, the time blocking approximation (TBA) is used.
It is shown that the TBA+ version of the model is quite efficient for eliminating
the coupling of the SDM with the physical modes.
However, the difference between
the results of the realistic calculations of the $E1$ excitations within
the TBA and TBA+ is very small if the effective external-field $E1$ operators are used.
The reason is that the strength of the SDM fragments in the TBA response function
is strongly suppressed by the effective operators.
At the same time, the difference between the TBA and TBA+ results
becomes appreciable in the case of the bare $E1$ operators.

\begin{acknowledgements}
This work was supported by the Russian Foundation for Basic Research,
project number 21-52-12035. The research was carried out using
computational resources provided by the Computer Center of
St. Petersburg State University.
\end{acknowledgements}

\bibliographystyle{apsrev4-1}
\bibliography{zet}
\end{document}